\documentclass[aps,prl,showpacs,twocolumn]{revtex4-1}

\usepackage{graphicx}
\usepackage{dcolumn}
\usepackage{bm}
\usepackage{amsmath,amssymb}

\usepackage{color}

\begin{document}

\title{The Influence of Network Topology on Sound Propagation in Granular Materials}

\author{Danielle S. Bassett$^1$\footnote{Corresponding author. Email address: dbassett@physics.ucsb.edu},
Eli T. Owens$^2$, Karen E. Daniels$^2$, Mason A. Porter$^{3,4}$}

\affiliation{$^1$Department of Physics, University of California, Santa Barbara, CA 93106, USA,
$^2$Department of Physics, North Carolina State University, Raleigh, NC 27607, USA,
$^3$Oxford Centre for Industrial and Applied Mathematics, Mathematical Institute, University of Oxford, Oxford OX1 3LB, UK,
$^4$CABDyN Complexity Centre, University of Oxford, Oxford, OX1 1HP, UK}

\date{\today}

\begin{abstract}
Granular materials, whose features range from the particle scale to the force-chain scale to
the bulk scale, are usually modeled as either particulate or continuum materials. In contrast with
either of these approaches, network representations are natural for the simultaneous examination of
microscopic, mesoscopic, and macroscopic features. In this paper, we treat granular materials as
spatially-embedded networks in which the nodes (particles) are connected by
weighted edges obtained from contact forces.  We test a variety of network measures for their utility
in helping to describe sound propagation in granular networks and find that network diagnostics can
be used to probe particle-, curve-, domain-, and system-scale structures in granular media.
In particular, diagnostics of meso-scale network structure are reproducible across experiments,
are correlated with sound propagation in this medium, and can be used to identify potentially
interesting size scales.  We also demonstrate that the sensitivity of network diagnostics
depends on the phase of sound propagation. In the injection phase, the signal propagates
systemically, as indicated by correlations with the network diagnostic of global efficiency.
In the scattering phase, however, the signal is better predicted by meso-scale community structure,
suggesting that the acoustic signal scatters over local geographic neighborhoods.
Collectively, our results demonstrate how the
force network of a granular system is imprinted on transmitted waves.
\end{abstract}

\pacs{
64.60.aq, 
43.25.+y, 
81.05.Rm   
}

\keywords{spatial networks, community structure, granular materials}

\maketitle


During the past 15 years, techniques from areas of physics such as
statistical mechanics and nonlinear dynamics have been used to make important
advances in studying networks across myriad disciplines \cite{newman2010}.
Conversely, the perspective of networks can also
play important roles in physical problems, as there is a large
class of heterogeneous systems such as foams, emulsions, and granular
materials \cite{VanHecke2010, Liu-2010-JTM} for which the
connectivity of the constituent elements is an important factor in the
deviation of their behavior from continuum models. In fact, the discontinuous
nature of granular materials led to the early idea of a fabric structure governing the anisotropic behavior of such materials \cite{Oda1972, Oda1980, Oda1982}.

\begin{figure}
\includegraphics[width=.8\columnwidth]{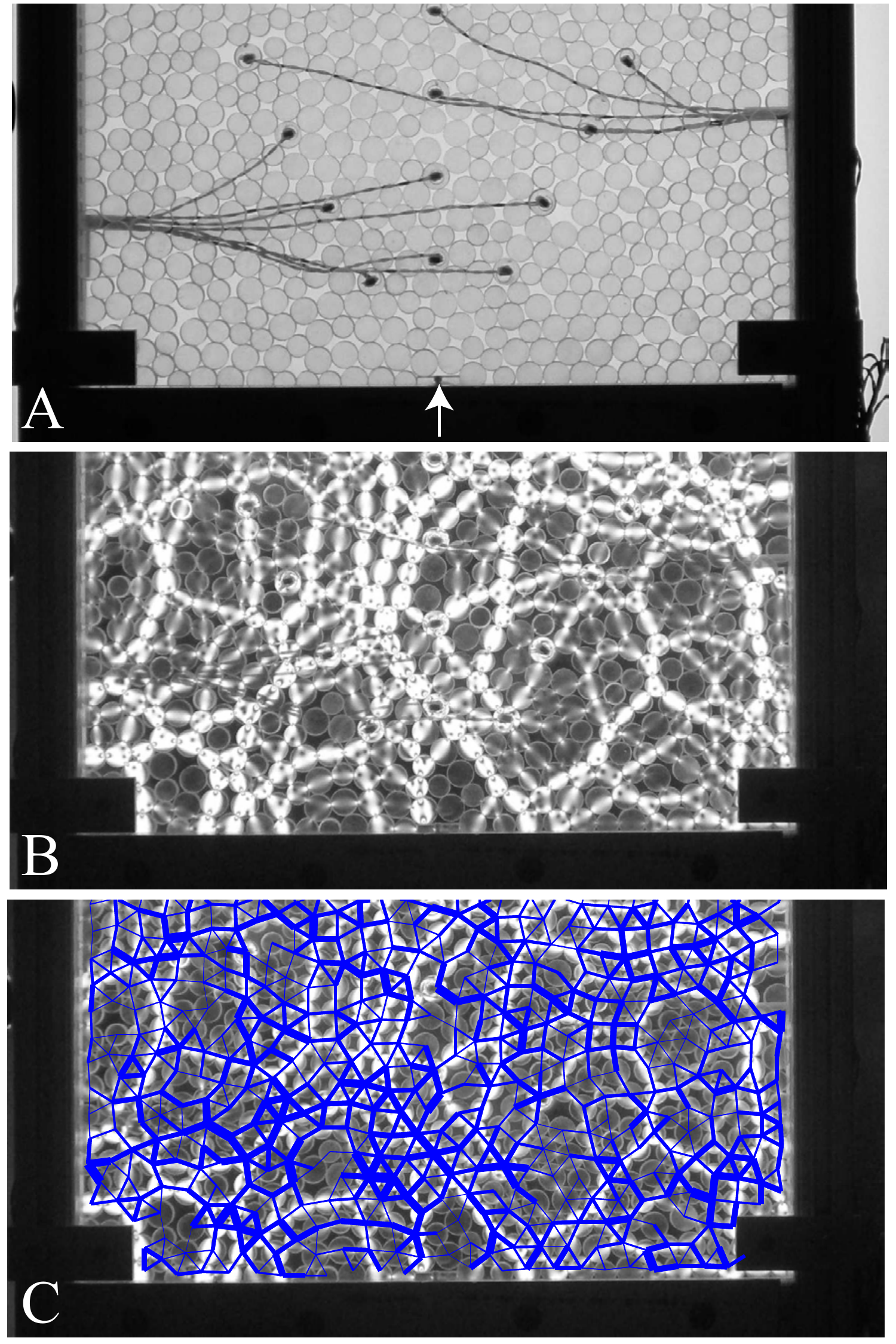}
\caption{\label{fig:exp} [Color online]
(A) Image of a 2D vertical aggregate of photoelastic disks confined in a single layer. The driver
position is marked with an arrow. Several particles are embedded with a piezoelectric sensor, for which wires are visible.
(B) The internal stress pattern within the photoelastic particles manifests as a network of force chains.
(C) Blue lines show a weighted graph, which is determined from image processing and overlayed on image (B).
An edge between two particles (nodes) exists if the two particles are in physical
contact with each other; the forces between particles give the weights of the edges.}
\end{figure}

We investigate whether studying the rich and complex dynamics of granular materials \cite{Jaeger1996} using network analysis can provide new insights into the underlying physics. This treatment is a natural one, because granular materials can be represented as spatially-embedded networks \cite{Barthelemy2010} composed of nodes (particles) and edges (contacts between particles) with definite locations in Euclidean space \cite{Herrera2011, Walker-2010-TED}. In Fig.~\ref{fig:exp}, we show a quasi-two-dimensional (quasi-2D) granular system composed of photoelastic disks that permits the determination of both the \emph{contact network} and the
interparticle forces. The forces between particles in these systems are
non-homogeneous, and they form a network of
chain-like structures that span the system (see Fig.~\ref{fig:exp}B). This \emph{force chain
network} has the same topology as the contact network but contains edges that are weighted by the inter-particle forces (Fig.~\ref{fig:exp}C). This is exciting from a networks perspective, as it allows us to study the influence of network topology on `network geometry' in a spatially-embedded system. From the perspective of granular materials, earlier work
suggests that force chains provide the main supporting structure for
static and dynamic loading \cite{Liu-1995-FFB, Howell1999}.

Because signal propagation in granular and heterogeneous materials \cite{Nesterenko2001} is of considerable importance to numerous industrial and natural systems, it has been the topic of many investigations.
A longstanding question is how to reconcile the failure of continuum models of granular sound
propagation \cite{Digby1981, Goddard1990, Velicky2002, Somfai2005}, as such
models fail to quantitatively describe important heterogeneous and nonlinear
features of acoustic speed \cite{Liu1994, Jia1999, Makse1999, Makse2004}. The
presence of force chains has been suggested as a potential confounding
phenomenon that might underlie the failure of previous physical models of
sound propagation \cite{Liu1994, Owens2011}. Ultimately, it would be beneficial to quantify how the pressure or strain
state of a system is imprinted on transmitted waves, and to understand how to
use these waves to accurately detect buried objects or reservoirs of oil.

An increasing body of work has used tools from areas like network science and computational homology to
obtain insights on the structural properties of granular materials \cite{Herrera2011, Walker-2010-TED, konst2012}
and other continuous media \cite{zohar2011}.
Indeed, a networks perspective provides a valuable complement to the standard ways of studying granular materials.
In the present paper, we analyze experimental data using network analysis to investigate the role of force-weighted
 contact networks in sound propagation. The use of photoelastic particles combined with high-speed imaging allows
us to gain insight into internal
force structures and particle-scale sound propagation that are not readily
available in ordinary granular materials.  We find that geographic community
structure provides a fundamental constraint to sound propagation,
illustrating that contact topology alone is insufficient to understand signal
propagation in granular materials.


\section{Experiments}

We perform experiments on a vertical 2D granular system of bidisperse disks
confined between two sheets of Plexiglass,
which have been slightly lubricated with baking powder to reduce friction with the container walls.
The top of the container is open and the particles are confined exclusively by gravity. The particles are $6.35$~mm thick and have diameters $d_1 = 9$~mm and $d_2=11$~mm, and are cut from Vishay PSM-4 photoelastic material to provide measurements of the internal forces.  We show example images in Fig.~\ref{fig:exp}. These particles have an elastic modulus of $E=4$~MPa, and they are sufficiently dissipative that propagating sound waves experience an approximately exponential decay as a function of distance from the source. The use of such soft, dissipative particles differs from previous work~\cite{Lazaridi1985, Coste1997, Job2005, Boechler2010, Huillard2011a}, where much harder particles have been used. Further details about the apparatus are described in Ref.~\cite{Owens2011}.

We excite acoustic waves from the bottom of the system by sending pulses of five
$750$~Hz sinusoidal waves with a voice coil driver attached to a $20$~mm
wide platform; maximum particle displacements are on the order of $5$~$\mu$m.
To assess the reliability of network diagnostics, we repeat the experiments
for $17$ different particle configurations, each of which is obtained by manual rearrangement. We restrict our analysis to a region of the system that contains just over $N=400$ particles.
This subsystem corresponds to a region in which vertical force gradients are minimized due
to the Janssen effect  \cite{Janssen1895}.

We compute particle positions and forces using two high-resolution pictures
of the static system and one high-speed movie that captures the system
dynamics. We took one static image without the polarizer (see Fig.~\ref{fig:exp}A) and used it to determine particle positions and contacts
 \cite{Majmudar2007, Owens2011}.  We take a second static image using
polarizers (see Fig.~\ref{fig:exp}B), and we use this together
with the contact locations to estimate the normal forces at each contact
using the methodology described in Ref.~\cite{Owens2011}. We measure the
amplitude and location of sound propagation using a high-speed camera
operating at $4000$~Hz; the camera records $80$ frames of data (20~ms)
containing both the injection of the signal ($0<t<40$) and its dissipation
($40<t<80$). For each particle in each frame, we compute $\Delta
I(x,y,t)=I(x,y,t)-I(x,y,t_0)$, which measures how much the brightness $I$ of
the particle changes with respect to its unperturbed brightness. In earlier work \cite{Owens2011}, we used piezoelectric sensors embedded in a subset of the particles to determine that $\Delta I$ is proportional to the change in stress on that particle. Using $\Delta I$ allows us to follow the propagating signal through all particles in the measurement region.

To determine which particles are in contact, we use the positions of
the particle centers, which are determined from the static image of the
system using a Hough transform. If the distance between two particle centers
is less than $1.05$ times the sum of their radii, we treat the particles as
being in contact. This method overcounts the number of true contacts.
However, the effect of such overcounting is minimized by the fact that false
contacts are assigned a force value of almost zero when we apply our image-processing
techniques to find the contact forces. Accordingly, they do not contribute to the structure of the weighted network.

For each experimental run, we construct both an unweighted (binary) and a
weighted network, which correspond respectively to an underlying contact
network and a force-chain network (see Fig.~\ref{fig:exp}).  In each type of network, the nodes
represent the particles in the system. In the binary network ${\bf A}$, an
edge exists between node $i$ and $j$ (i.e., $A_{ij} = 1$) if node $i$ is in
contact with node $j$; otherwise, $A_{ij} = 0$.  The weighted network ${\bf
W}$ contains the same edges, but each element $W_{ij}$ now has a value that
is given by an estimate of the normal force $f_{ij}$ between particles $i$ and $j$, normalized by the mean force $\overline{f}$ of all contacts: $W_{ij} = {f_{ij}}/{\overline{f}}$.


\section{Results}

We assess the global organization of the networks using 21 candidate diagnostics for ${\bf A}$ and 8 candidate diagnostics for ${\bf W}$. We define each diagnostic in Appendix A, where we also include descriptions to provide intuition about what each of them measures, as well as their possible physical significance to the granular system that we study. We examine the reliability of these diagnostics across experimental runs in Appendix B, and we compare the binary-network diagnostics to those in a null model constructed using an ensemble of random geometric graphs (RGGs) \cite{penrosergg} in Appendix C. We examine 4 diagnostics (clustering coefficient, geodesic node betweenness, optimized modularity, and global efficiency) in further detail.  Each diagnostic can be defined for both binary and weighted networks, and each is helpful for obtaining insights into a particular type of spatial structure in the system: particles (cluster coefficient), curves (betweenness), meso-scale domains (via community structure determined from modularity optimization), and the entire system (global efficiency).  We describe our results in the sections below.


\subsection{Scale Sensitivity of Network Diagnostics}

\begin{figure}[]
\includegraphics[width=.47\textwidth]{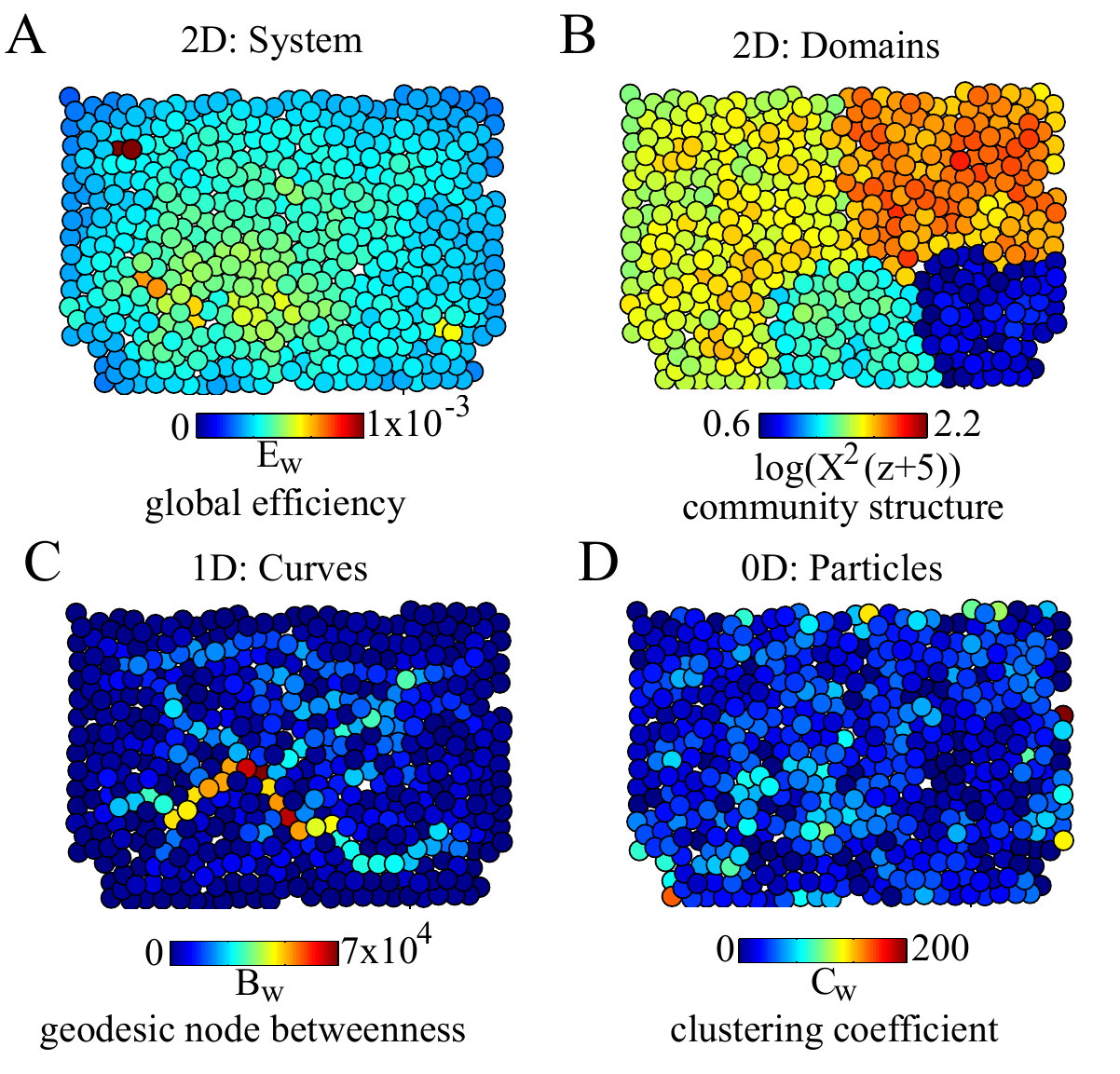}
\caption[]{\label{fig:scale} [Color online]
Example distributions of several network diagnostics for a sample granular packing. The network characteristics that we examine include
(A) global efficiency $E_{w}$,
(B) community structure, which we visualize using the quantity $X^2 (z+5)$, where $X$ is the community label,
(C) geodesic node betweenness $B_{w}$, and
(D) clustering coefficient $C_{w}$. This figure illustrates their respective sensitivities to system-scale, domain-scale, curve-scale, and particle-scale structure. The quantity $X^2 (z+5)$ allows us to visualize both the community label ($X$) and the intra-community strength z-score ($z$, see Eq.~\ref{eq:zscore}) simultaneously; we chose the constant $5$ purely for visual clarity.}
\end{figure}

A key advantage of using network tools to study granular materials
is that different network diagnostics (which we define and discuss in detail in Appendix A)
are sensitive to different system scales, and this is especially helpful for spatially-embedded
systems like granular packings (see Fig.~\ref{fig:scale}). Our results indicate that the global
efficiency $E_w$ [see Eqs.~(\ref{eq:Ewc}) and (\ref{eq:Eww})] is a system-level property with
smallest values along the perimeter of the system and largest values in the center.
Community structure and its associated community label $X$ [see Eqs.~(\ref{modmod}) and (\ref{eq:Xw})] and intra-community strength z-score [see Eqs.~\ref{eq:zscore}]
is a meso-scale property and can be used to probe intermediate structural features.
We find that geodesic node betweenness $B_w$ [see Eqs.~(\ref{eq:Bwc}) and (\ref{eq:Bww})]
can be thought of as a one dimensional property in these materials because it
is sensitive to curve-like structures in the
network.  Finally, we find that clustering coefficient $C_w$ [see Eqs.~\ref{eq:Cc} and \ref{eq:Cw})]
is sensitive to particle-scale features of the network.  In a later section, we report how each of
these network diagnostics correlates with sound
propagation through the granular material.

\subsection{Identifying a Characteristic Size Scale}

\begin{figure}
\begin{center}
\includegraphics[width=0.49\textwidth]{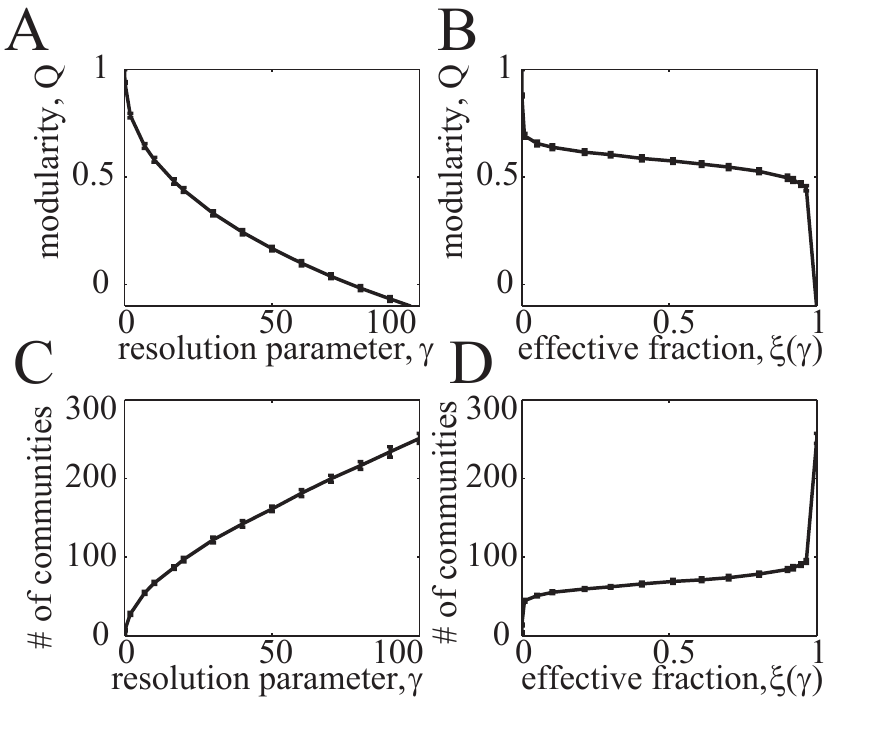}
\caption{
(A,C) Modularity index $Q_{w}$ and
(B,D) number of communities as a function of the resolution parameter
(A,B)  $\gamma$ and
(C,D) the effective fraction of antiferromagnetic edges $\xi(\gamma)$. Error bars indicate the standard deviation over the 17 experimental runs.
\label{fig:resolution}}
\end{center}
\end{figure}

An ongoing challenge in the study of granular systems is identifying and measuring characteristic size scales within granular materials, from the perspective of either particles or force chains \cite{Liu-2010-JTM, VanHecke2010, Ostojic2006}.
Network modularity provides a novel means to measure such size scales via the identification of community sizes. We find that the optimal value of modularity is a reliable diagnostic for the
structure of both the binary and weighted networks (see Appendix B). To seek characteristic community sizes, we also examine community structure as a function of a resolution parameter $\gamma$ \cite{Reichardt2006,Fortunato2007,Fortunato2010}.
The modularity index is
\cite{Porter2009}
\begin{equation} \label{eq:Qw}
	Q_{w}=\sum_{ij} [W_{ij}-\gamma P_{ij}] \delta(g_{i},g_{j})\,,
\end{equation}
where node $i$ is assigned to community $g_{i}$, node
$j$ is assigned to community $g_{j}$, $\delta(g_{i},g_{j})=1$ if $g_{i} =
g_{j}$ and it equals $0$ otherwise, and $P_{ij}$ is the expected weight of
the edge connecting node $i$ and node $j$ under a specified null model.  We
used the usual Newman-Girvan null model, in which the expected strength
distribution of the network is preserved but ends of edges are rewired
uniformly at random \cite{newmangirvan,Porter2009}. We employed the Louvain locally
greedy algorithm to optimize modularity \cite{Blondel2008}, and we varied
the resolution parameter $\gamma$ from $0.001$ to $100$. Low values of $\gamma$ probe large spatial scales, and high values probe small scales.  When we increase $\gamma$, the number of communities increases (as expected), and the modularity decreases.  See Fig.~\ref{fig:resolution}A,C.

One can think of the term $J_{ij}(\gamma) = W_{ij}-\gamma P_{ij}$ in Eq.~(\ref{eq:Qw})
as a particular choice of interaction strength between a pair of spins in a Potts
model \cite{Porter2009,Reichardt2004,Reichardt2006}. We exploit this analogy with the
Potts model to transform the resolution parameter $\gamma$ so that it measures the
effective fraction of
antiferromagnetic edges $\xi(\gamma)$ in a network \cite{Onnela2012}.
We define $l^{W}(\gamma)$ to be the number of negative elements of $\mathbf{J}(\gamma)$.
The transformed resolution parameter is then
\begin{equation}
	\xi(\gamma) = \frac{l^{W}(\gamma) - l^{W}(\Lambda_{\mathrm{min}})}{l^{W}(\Lambda_{\mathrm{max}}) - l^{W}(\Lambda_{\mathrm{min}})} \in [0,1] \,,
\end{equation}
where $\Lambda_{\mathrm{min}}$ is the largest number of negative entries
$J_{ij}(\gamma)$ for which an $N$-node network forms a single community and
$\Lambda_{\mathrm{max}}$ is the smallest number for which the network still
forms $N$ communities of size $1$.

We examine community structure as a function of the transformed resolution parameter $\xi(\gamma)$,
which we vary between $0$ and $1$. The optimized modularity and the number of communities both
change gradually for most of the
$\xi(\gamma)$ range (see Fig.~\ref{fig:resolution}B,D), although abrupt changes are
evident for very low and very high values of $\xi(\gamma)$. The gradual
change hints at an interesting size scale, which occurs in partitions that contain about $50-100$ communities (with a characteristic size of roughly $5-8$ particles). One possibility is that this size corresponds to the width of a shear band, which arise in a variety of materials with particulate structure \cite{Schall-2010-SBM}. Another possibility is that this size corresponds to the `cutting' length scale $\ell^*$ \cite{Wyart-2005-RAS}, which is set by a community size at which the excess (over-constrained) number of contacts in the bulk of a region is equal to the number of contacts around the perimeter. If this latter association is correct, then the mean number of particles per community
would scale with the confining pressure. Future experiments can test this hypothesis.


\subsection{Geography of Community Structures}

\begin{figure*}[]
\includegraphics[width=.9\textwidth]{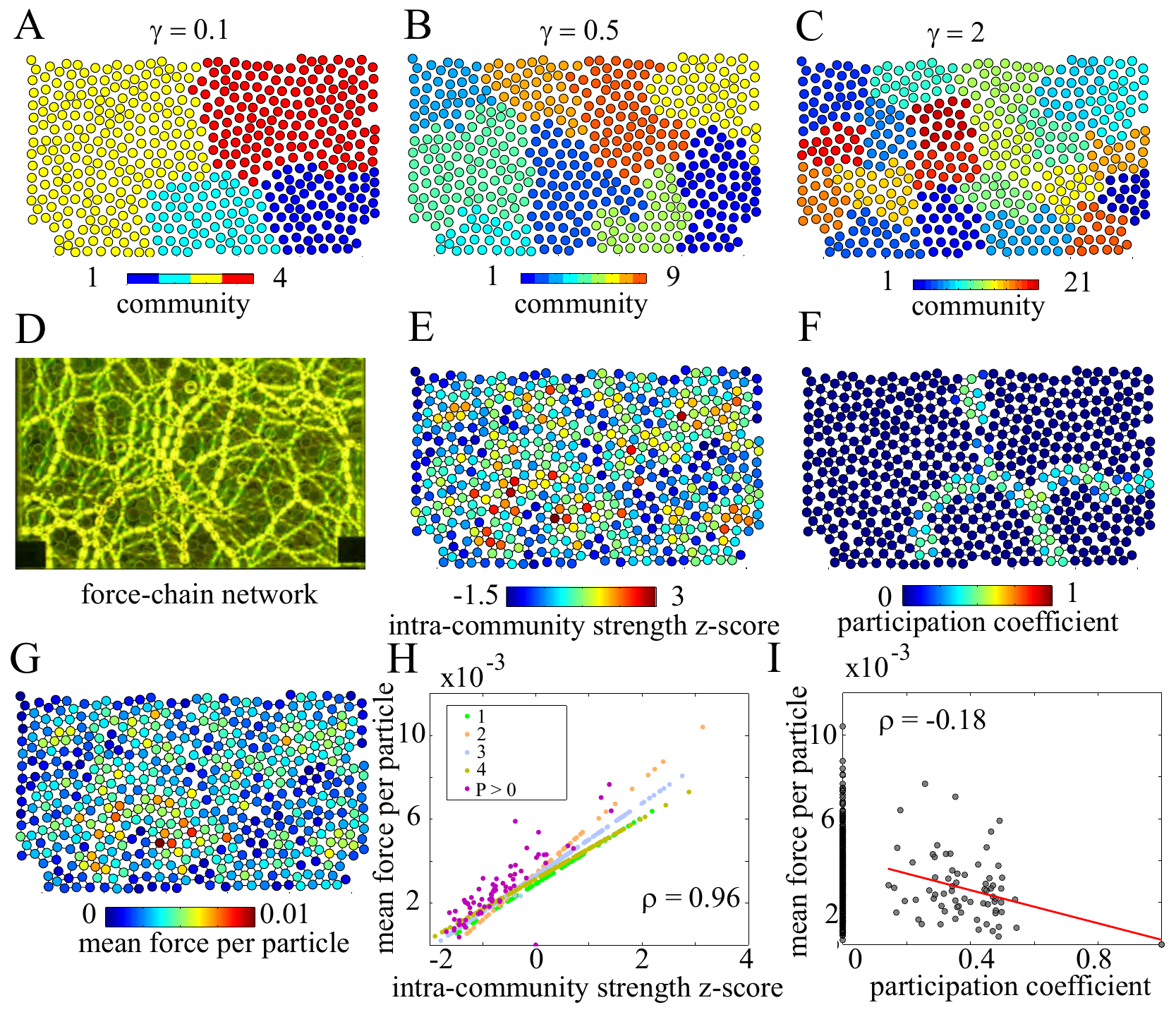}
 \caption[]{\label{fig:9panes} [Color online]
The geographic sizes of communities tend to
decrease as the resolution parameter $\gamma$ is increased from (A) low ($\gamma = 0.1$ to (C) high $\gamma = 1$ values.  We color particles according to their community label.
One can examine community structure of the (D) force-chain network  using
geographic location,
(E) intra-community strength z-score $z_i$,
(F) participation coefficient $P_i$, and
(G) mean force per particle  $S'_i$.
Example scatterplots for a single experimental run showing that mean force $S'_i$ on particle $i$ is
(H) positively correlated with the intra-community strength z-score (the Spearman correlation coefficient for this run is $\rho \approx 0.96$ and the p-value is $p \approx 1.9\times 10^{-283}$) and
(I) negatively correlated with the participation coefficient ($\rho \approx -0.18$ and $p \approx 1.8 \times 10^{-5}$). In panel (H), nodes assigned to communities 1 through 4 and whose participation coefficients are equal to 0 are displayed using different colored markers. Nodes in any of the 4 communities whose participation coefficients are greater than 0 (so-called `boundary nodes') are displayed using purple markers. The mean correlations for $z$ and $P$ over experimental runs and values of the resolution parameter are $\rho \approx 0.87\pm0.08$ and $\rho \approx -0.16\pm0.05$, respectively. We show the results for one experimental run ($\#2$) in this figure, and the results for the other runs are similar.
}
\end{figure*}

Using modularity optimization \cite{newmangirvan,Newman2006,Porter2009,Fortunato2010},
we find that the force-chain network exhibits geographically-constrained community structure: groups of particles in close spatial proximity are more likely to be a part of the same community (i.e., to contact one another with a large force) than particles that are farther
apart. We examine this local neighborhood structure over a variety of size
scales by varying the resolution parameter $\gamma$. We show representative results for large spatial scales in Fig.~\ref{fig:9panes}A-C. We also note that the communities that we identify in granular force networks resemble those in spatial entities like states or countries, whose borders are determined in part by physical boundaries between neighboring geographic domains.

Importantly, because the optimization of $Q$ is NP-hard \cite{brandes08}, one
does not expect an optimization algorithm to give a global optimum of $Q$.
Instead, we harness numerous near-degeneracies \cite{Good2010} among good local optima of $Q$ by estimating $Q$ 100 times.
We find that the these 100 values of $Q$ for a given run at a given $\gamma$
vary by approximately $1 \times 10^{-14}$, and the similarity in particle
assignments to communities is approximately $0.98$.  We quantify this using partition similarity  \cite{Kuncheva2004,Danon2005}, which ranges from $0$
(not similar at all) to $1$ (identical). These results indicate that the local
geographic structures that we are identifying in the 2D granular system are robust, suggesting the potential for identifying reproducible 2D `geographic' regions.

To probe the role of each particle in the community structure of a force-chain network (see Fig.~\ref{fig:9panes}D), we use the intra-community strength
z-score $z_i$ to measure how well connected a node is to other particles in
its community and the participation coefficient $P_i$ to measure how the
connections emanating from a particle are spread among particles in the different communities \cite{guimera05}.

The intra-community strength z-score is
\begin{equation}
 	z_{i} = \frac{S_{g_{i}} - \bar{S}_{g_{i}}}{\sigma_{\bar{S}_{g_{i}}}}\,,
\label{eq:zscore}
\end{equation}
where $S_{g_{i}}$ denotes the strength (i.e., total edge weight) of node $i$'s edges to other nodes in its own community $g_{i}$, the quantity $\bar{S}_{g_{i}}$ is the mean of $S_{g_{i}}$ over all of the nodes in $g_{i}$, and $\sigma_{S_{g_{i}}}$ is the standard deviation of $S_{g_{i}}$ in
$g_{i}$. The strength of node $i$ is denoted by $S_i$ and gives the total force of all contacts on particle $i$.

The participation coefficient is \cite{guimera05}
\begin{equation}
	P_{i} = 1 - \sum_{g=1}^{N_{m}} \left(\frac{S_{ig}}{S_{i}}\right)^{2}\,.
\end{equation}
where $S_{ig}$ is the strength of edges of node $i$ to nodes in community $g$ \cite{guimera05}.

In Fig.~\ref{fig:9panes}E-F, we show the intra-community strength z-score and participation coefficient for the community structure depicted in Fig.~\ref{fig:9panes}A. Particles that are geographically central to a community tend to have higher values of $z_i$ and lower values of $P_i$ than particles at the geographic periphery of a community. From a physical perspective, $z_i$ tends to be highest in
particles with many force chains passing through them (compare, e.g., Figs.~\ref{fig:9panes}D and \ref{fig:9panes}G) and high values of $P_i$ are associated with the boundaries between communities (where there are few force chains).

We test whether the observed properties of community structure in our granular systems are related statistically to the inter-particle forces that constitute the
force-chain structure (see Fig.~\ref{fig:9panes}D,G) by examining the relationship
between intra-community strength z-score $z_i$, participation coefficient
$P_i$, the normalized node strength $S'_{i} = S_{i}/N$, (i.e., the mean force of all
edges emanating from a node) and the amplitude of the acoustic signal $\Delta
I$ using the Spearman rank correlation coefficient $\rho$, which is defined
as the Pearson correlation coefficient between ranked variables. We use the
Spearman coefficient rather than the Pearson coefficient due to the
non-normal distributions of $\Delta I$ values over particles.

We find that the mean force of all contacts on a particle (i.e., $S'_{i}$) is
significantly positively correlated with $z_i$ (see Fig.~\ref{fig:9panes}H). The mean
value of the Spearman rank correlation coefficient $\rho$ over experimental
runs and resolution-parameter values is $\rho \approx 0.71\pm0.02$ (where $0.02$ gives the standard deviation over experimental runs). This strong positive correlation indicates
that particles at the centers of communities are likely to have more or
stronger force chains running through them. We also find that $S'_{i}$ is
negatively correlated with $P$ (Fig.~\ref{fig:9panes}I).
The mean $\rho$ over experimental runs and values of the resolution parameter is $\rho \approx
-0.05\pm0.04$, where we again take the standard deviation only over experimental runs. This negative correlation indicates that inter-community boundaries occur at particles with
fewer or weaker force chains. Note additionally that a large fraction of the particles have $P=0$. This is a consequence of the fact that the communities are geographically constrained such that the majority of particles have contacts only within their own community.

The relationship between $z$, $P$, and $S'$ is expected mathematically. For example, if the edges of node $i$ all lie within its own community, then $S'$ and $z$ are related linearly according to the following equation: $S_{i} = z_i \times \sigma_{S_{g_{i}}} + \bar{S}_{g_{i}}$, where $S_{g_{i}}$ is the strength of edges of node $i$ to other nodes in its
community $g_{i}$, and $\bar{S}_{g_{i}}$ is the mean of $S_{g_{i}}$ over all
of the nodes in $g_{i}$.  This linear relationship is evident for the four
communities that we show in Fig.~\ref{fig:9panes}H.  Nodes whose connections span
more than one community (so-called `boundary nodes', for which the value of
$P$ is greater than $0$) are not so simply related.


\subsection{Signal Propagation on Force-Weighted Contact Networks}

\begin{figure}[]
\includegraphics[width=.5\textwidth]{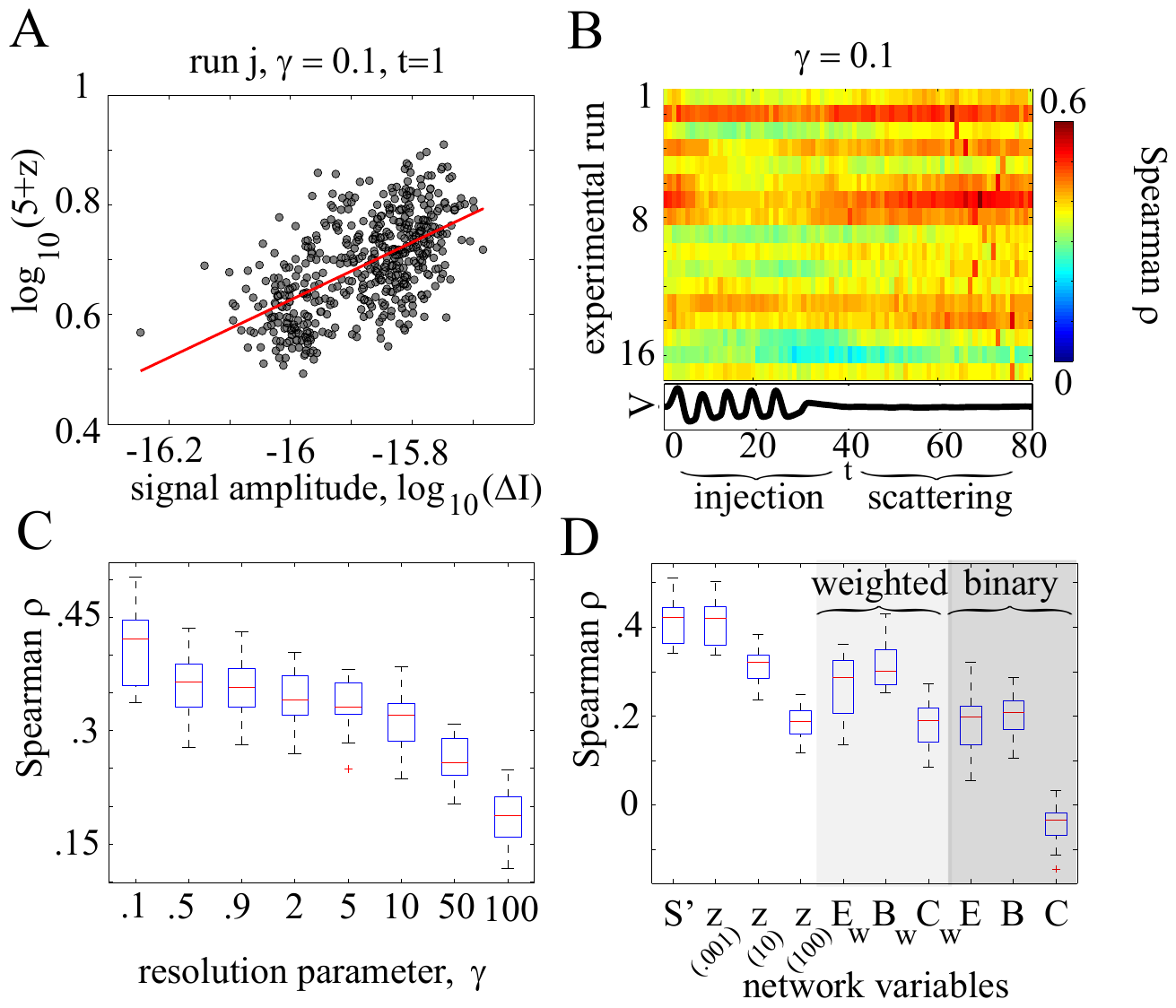}
\caption[]{\label{fig:sound} [Color online]
(A) An example scatterplot between logarithms of the intra-community strength z-score [$\log_{10}(z+5)$] and the amplitude of the acoustic signal [$\log_{10}(\Delta I$] at $t=1$ for a single experimental run ($j = 2$) and
resolution-parameter value ($\gamma = 0.1$). The constant $5$ was added to $z$ to ensure that all values were positive prior to taking the logarithm. The Spearman rank correlation
coefficient is $\rho \approx 0.57$ and the $p$-value is $p \approx 2.1 \times
10^{-45}$.
(B) Correlation between $\mathrm{log_{10}(z+5)}$ and
$\mathrm{log_{10}(\Delta I)}$ for all 17 runs as a function of time: $t<40$
is the acoustic signal injection phase, and $t>40$ is the
acoustic signal scattering phase. In the bottom part of panel (B), we show a trace of the voltage $V$ of a piezo particle; the injected signal has an amplitude of 0.5V.
(C) Correlation between $\mathrm{log_{10}(z+5)}$ and $\mathrm{log_{10}(\Delta I)}$
as a function of $\gamma$, where $\gamma \in [0,2]$ is increased in
increments of 0.1. We averaged the correlation over all 80 time points at
which $\Delta I$ was measured, and the mean $\rho$ over runs $\gamma$ values,
and time was $0.34\pm0.06$. Box plots show the variability over experimental
runs.
(D) Correlation between $\mathrm{log_{10}(\Delta I)}$ and a
variety of network diagnostics: intra-community strength z-score $z = z(\gamma)$ for
$\gamma=0.001$, $\gamma=10$, and
$\gamma=100$; and weighted (white background, left of the
figure) and unweighted (light gray background, middle of the figure) versions
of global efficiency [$E_{w}(i)$ and $E(i)$], geodesic node betweenness
[$B_{w}(i)$ and $B(i)$], and clustering coefficient [$C_{w}(i)$ and
$C(i)$]. For completeness, we also show results for the mean force $S'$ (left
of the figure), which was the variable previously reported to be correlated
with $\Delta I$  \cite{Owens2011}.
}
\end{figure}

Previous work in Ref.~\cite{Owens2011} has shown that the propagation of acoustic
signals is facilitated along strong force chains in granular materials, via the increased contact area at strong contacts. With this in mind, we test whether the geographic community structure of force-chain networks is
related to signal propagation. As the example shown in Fig.~\ref{fig:sound}A
indicates, we found that $z$, which we measured over a range of size scales
associated with resolution-parameter values $\gamma \in [0.001,100]$, is
significantly correlated with the signal amplitude $\Delta I$. (For this
example run, $\rho \approx 0.57$ and the p-value is $p \approx 2.1\times 10^{-45}$.)
The statistical correlation between network structure and signal
amplitude exists not only in the highly heterogeneous signal injection phase,
in which sound propagates from the driver to nearby particles, but also in the
more homogeneous scattering phase, in which sound reverberates throughout the
system. In Fig.~\ref{fig:sound}B, we show the results at $\gamma=0.1$; Figure~\ref{fig:sound}C shows that the mean $\rho$ over runs, $\gamma$ values, and time is
$0.34\pm0.06$. This suggests that similar dynamic principles underlie sound propagation in both injection and scattering phases.
We discuss the dynamics within the two phases in more detail in the next section.

In quantifying the relationship between the intra-community strength z-score
$z$ (a property of the algorithmically computed community structure) and the signal
propagation amplitude $\Delta I$, we note that the correlation between these
two variables is decreases as $\gamma$ increases (see Fig.~\ref{fig:sound}C). The
strength of the relationship between network structure and signal propagation
for small $\gamma$ suggests that partitions with a few large communities achieve
better estimates of the propagation behavior.
Indeed, as demonstrated in Fig.~\ref{fig:sound}D, when the network forms a single
community (for very low $\gamma$ values), the correlation between $z$ and
$\Delta I$ is similar to that between the mean force per particle ($S'$) and
$\Delta I$.

The retention of a correlation between $z$ and $\Delta I$ for larger values
of $\gamma$, for which the network is partitioned into more (and smaller) communities, stems
from the strong correlation between intra-community strength z-score and the
mean force (normalized strength) of a particle (see Fig.\ref{fig:9panes}H), the
latter of which is a particle-scale measurement and is independent of spatial
resolution. The relationship between the meso-scale (community structure) and
particle-scale (mean force on a particle, which is equal to a node's normalized strength) network
properties stems from the physical embedding of the granular system in
$\mathbb{R}^{2}$. A particle that is located geographically inside of a community has
all of its connections to other particles in its community because it
is constrained to connect only to its geographic neighbors (i.e., there can be no long-range contacts).
This is unlike most investigated real-world
networks \cite{Porter2009,Fortunato2010}, in which communities tend to be highly interconnected and most nodes have at least some connections to nodes in other communities.

To assess whether community structure is unique in its ability to predict
signal amplitude, we also examine other weighted network diagnostics that
are sensitive to different system dimensionalities (see Fig.~\ref{fig:scale}).
Our results suggest that community structure (see Fig.~\ref{fig:scale}B)
is a better predictor of signal propagation than system-scale
(e.g., global efficiency; see Fig.~\ref{fig:scale}A), curve-scale (e.g., geodesic
node betweenness; see Fig.~\ref{fig:scale}C), and particle-scale (e.g., clustering
coefficient; see Fig.~\ref{fig:scale}D) network diagnostics. See Appendix A for mathematical definitions and intuitive descriptions. In particular, clustering coefficient and $\Delta I$ are not strongly correlated, so
triangles of contacts do not appear to be important for signal propagation. See the
comparison in Fig.~\ref{fig:sound}D.


\subsection{Phase Sensitivity of Network Diagnostics}

\begin{figure}[]
\includegraphics[width=.8\columnwidth]{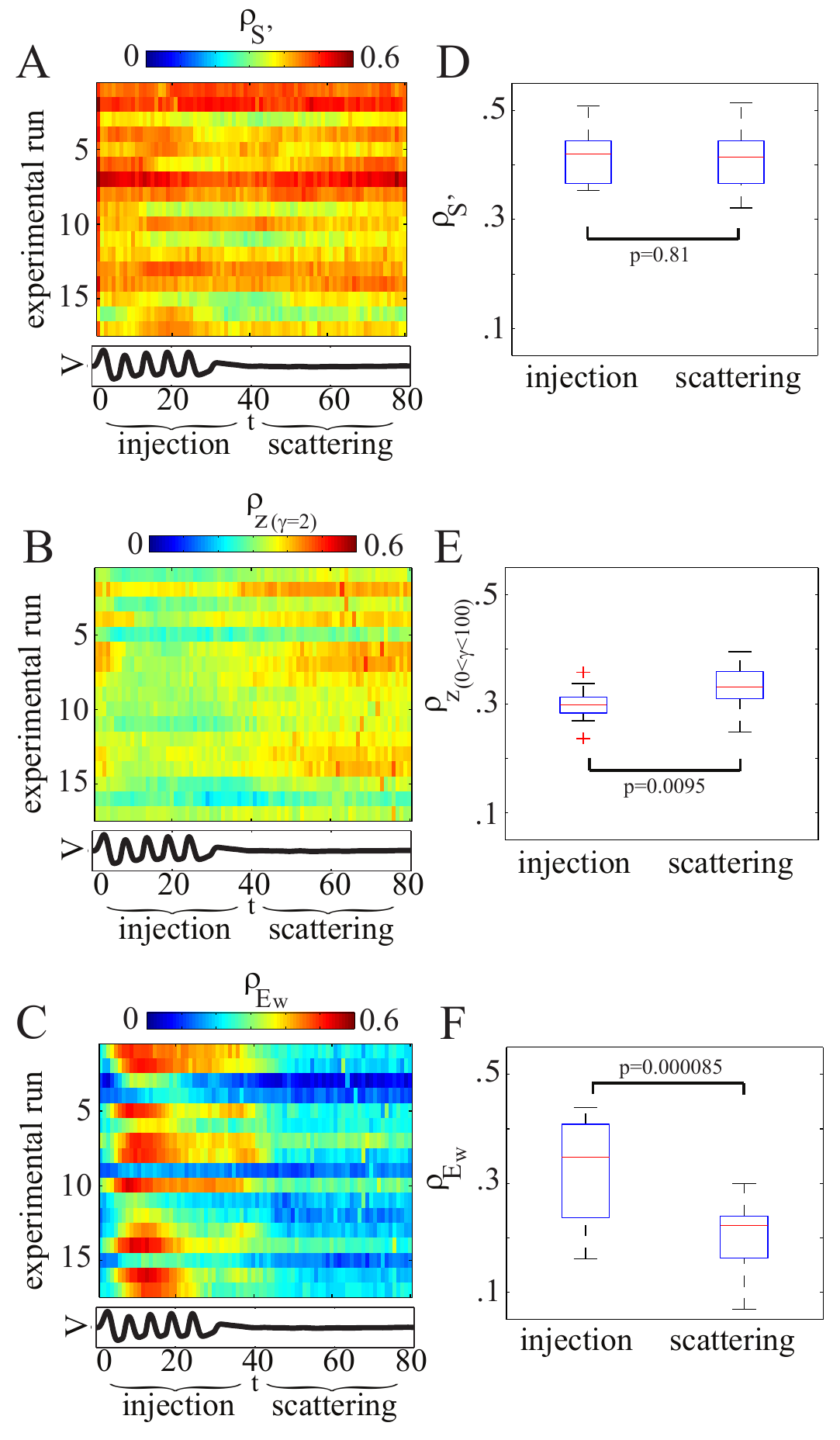}
\caption[]{\label{fig:phase} [Color online]
Spearman correlations between signal amplitude $\Delta I$ and
(A) the mean force per particle $S'$,
(B) the intra-community strength z-score$z$ for $\gamma=2$, and
(C) the global efficiency $E_{w}$.
Data is  for all experimental runs and
for all times, including both injection (left) and scattering (right) phases.
Box plots of the Spearman correlations between signal amplitude $\Delta I$ and the
three diagnostics shown in (A-C):
(D) mean force per particle ($\rho_{S'}$),
(E) intra-community strength z-score $z$ ($\rho_{z}$), and
(F) global efficiency ($\rho_{E_{w}}$).
We have averaged the correlations that we show in the box plots over the injection (left) and scattering (right)
phases. For (E), note that we also average the correlations over the
resolution parameter $\gamma$. Using MATLAB notation, the precise values of $\gamma$ that we considered are $[0.001:0.001:0.009,~ 0.01:0.1:1,~ 2:3,~ 4:0.1:20,~ 30:10:100,~ 200:100:1000]$. The reported p-values indicate the results of 2-sample
t-tests.
}
\end{figure}

Although the correlation between $z$ and signal amplitude
is strong in both injection and scattering phases for small $\gamma$ (i.e., large
community size), it is higher in the scattering phase than in the
injection phase when averaged over all resolutions ($\gamma\in [0.001,100]$; see Fig.~\ref{fig:phase}B,E) 
We do not observe such sensitivity to phase for the mean force per
particle (see Fig.~\ref{fig:phase}A,D).
Interestingly, the signal propagation during the injection phase is more strongly correlated with the
global efficiency than it is during the scattering phase (see Fig.~\ref{fig:phase}C,F), suggesting that the acoustic signal propagates over the shortest weighted paths during the injection phase.  These results illustrate insights from network analysis that one cannot obtain from particle-scale measurements: signal propagation
during injection is well characterized by shortest paths that span the
system, whereas it is characterized by local neighborhood structure during scattering.
An interesting question is whether the amplitude of the injected signal affects the
size of the geographic neighborhood through it propagates.

\begin{figure}[]
\includegraphics[width=.49\textwidth]{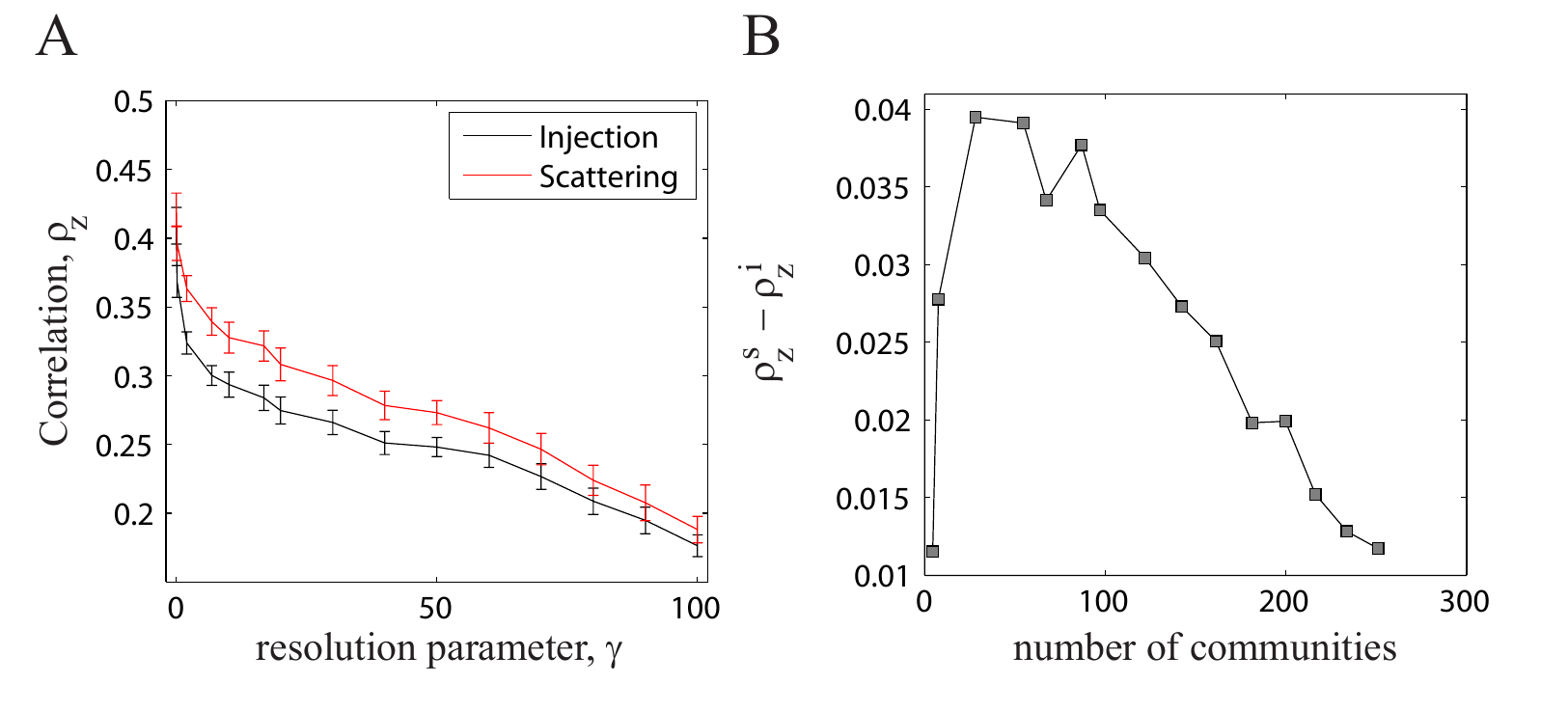}
\caption[]{\label{fig:phase-res} [Color online]
(A) Spearman correlations between signal amplitude $\Delta I$ and
the intra-community strength z-score $z$ averaged over the injection (black)
and scattering (red) phases as a function of the resolution parameter $\gamma$.
(B) Difference between the correlation in the scattering ($\rho_{z}^{s}$) and
injection ($\rho_{z}^{i}$) phases as a function of the number of communities in the partition.
We find the greatest sensitivity to phase for partitions with approximately
$50-100$ communities (i.e., for communities containing roughly $5-8$ particles), which corresponds to the size scale that identified as potentially interesting using the transformed resolution parameter.
}
\end{figure}

We also examine the sensitivity of the relationship between $z$ and $\Delta I$ to the
injection and scattering phases as a function of the resolution parameter
(see Fig.~\ref{fig:phase-res}A). The correlation between $z$ and signal
amplitude is consistently higher in the scattering phase than in the injection
phase throughout $\gamma \in [0.001,100]$. Furthermore, the
largest difference in the Spearman correlation between $z$ and $\Delta I$ for the scattering versus injection phases occurs for partitions with approximately $50-100$ communities,
corresponding to community sizes of roughly $5-8$ particles (see Fig.~\ref{fig:phase-res}B). This is similar to the size scale that we identified previously when using the transformed resolution parameter.

\section{Discussion}

A networks perspective provides a useful framework in which to study the material and dynamic properties of granular materials. Network diagnostics vary in their sensitivity to scales of the granular system: the particle-scale can be probed with a clustering coefficient, the curve-scale can be probed with geodesic node betweenness, the domain-scale can be probed with community structure, and the system-scale can be probed with global efficiency. Moreover, one can identify potentially interesting length/size scales in the system using meso-scale network features such as community structure.  As we show in Appendix C, one can also obtain physical insights into the geographic organization of the material by comparing the features of the actual networks to a null model consisting of an ensemble of random geometric graphs.

The dynamics of signal propagation on a network are best characterized by weighted diagnostics derived from the granular force-chain network, suggesting that the topology of the underlying (unweighted) contact network alone is not sufficient to explain signal propagation. In other words, one must also consider network geometry. This result underscores the important relationship between signal propagation and force-chain organization (see Fig.~\ref{fig:sound}D). Similar phenomena are likely relevant for a variety of energy transport problems (e.g., in sound, heat, and electricity) in a broad class of amorphous materials.

The algorithmic detection of communities is particularly useful in quantifying the effect of meso-scale network structure on signal propagation. We find that community structure is a better predictor of signal amplitude over the range of propagation phases than system-scale, curve-scale, or particle-scale measurements. Furthermore, by contrasting signal behavior during injection and scattering phases, we are able differentiate the sensitivities of system-wide and meso-scale network structure to sound propagation. Community structure seems to be a better predictor of signal propagation in the scattering phase than in the injection phase, suggesting that the sound scatters in local geographic neighborhoods. However, global efficiency predicts signal propagation better in the injection phase than in the scattering phase, suggesting a more system-wide dynamic distribution of sound.

Studying community structure allows one to investigate the meso-scale
architecture of the granular packing. In addition to the intra-community strength z-score
predicting particle sound amplitude, the identification of geographic communities provides a
quantifiable size scale, which may be useful for seeking the diverging length scale that is expected
at the jamming transition \cite{VanHecke2010, Liu-2010-JTM}.
The meso-scale nature of the sound propagation might be related to other meso-scale phenomena in granular physics, such as the spatial eigenmodes for soft (low-energy) modes, which are observed in simulations to take the form of localized swirls \cite{Wyart2005, Henkes2010c}.
Our approach also provides a framework to relate 1D structures (force chains) to 2D structures (geographic
domains).  It therefore might also prove useful in other settings --- such as
in the study of crystalline solids, where domain structure is critical to
system function.

The presence of correlated regions such as geographic communities in a granular material is
reminiscent of shear transformation zones (STZs \cite{Falk1998}), in which
localized regions throughout a sheared material have a higher propensity to
deform under shear.  Importantly, however, the community structure that we
compute spans the system, whereas STZs are relatively small structures
dispersed throughout system.  Also of interest is a comparison with the results of Ref.~\cite{Manning2011}, which illustrated that vibrational modes can identify soft spots in sheared systems.

In conclusion, using network analysis to study granular materials can be extremely
useful, as it can help characterize particle, curve, domain, and system-scale properties of such materials. In particular, the algorithmic detection of communities provides a means to identify potentially interesting characteristic size scales in such systems.
When combined with time-resolved acoustic measurements \cite{Owens2011}, such a networks perspective can
illuminate the meso-scale structures within which sound travels preferentially.
We found that particles that are well connected to
their community have larger-amplitude signals passing through them. Our results also suggest that signals scatter in local geographic neighborhoods but propagate more systemically during signal injection.
Investigation of both weighted and unweighted networks demonstrates that a
weighted network is a better predictor of sound propagation, suggesting
that the force-chain structure of the granular material is an important component
in sound propagation.  Our results demonstrate that one cannot examine only
system-scale or local-scale network features to understand how sound travels
through a granular material.  Importantly, one achieves a better description
of sound propagation when one includes how the particles relate to their
neighbors in a network.


\bigskip

\textbf{Acknowledgments} We thank Lee C. Bassett for helpful insights and
Jean Carlson, Aaron Clauset, Wolfgang Losert, Peter Mucha, Colin
McDiarmid, Zohar Nussinov, and Marta Sarzynska for useful comments. D.S.B. was supported by the David and Lucile Packard Foundation, Public Health Service Grant NS44393, the Institute for
Collaborative Biotechnologies through Contract W911NF-09-D-0001 from the US
Army Research Office, and the National Science Foundation (Division of
Mathematical Sciences-0645369). E.T.O and K.E.D were supported by a NSF
CAREER award DMR-0644743.  M.A.P. acknowledges the Statistical and Applied
Mathematical Sciences Institute, the Kavli Institute for Theoretical Physics,
and a research award (\#220020177) from the James S. McDonnell Foundation.


\section{Appendix A: Definitions of Network Diagnostics}

\subsection*{Diagnostics Applied to Unweighted Contact Networks}

To characterize structure of the binary (contact) networks, we examined 21
diagnostics: number of nodes, number of edges, global efficiency
 \cite{Latora2001}, geodesic node betweenness
centrality \cite{Freeman1977}\footnote{Betweenness centrality is often called
simply `betweenness', so we will do this in the remainder of this
supplement.}, random-walk node betweenness \cite{Newman2005}, geodesic edge
betweenness  \cite{Girvan2002}, eigenvector centrality \cite{Bonacich1972},
closeness centrality \cite{Sabidussi1966}, subgraph
centrality \cite{Estrada2005}, communicability \cite{Estrada2008}, clustering
coefficient \cite{Watts1998}, local efficiency \cite{Latora2001}, modularity
optimized using two different
algorithms \cite{Newman2006,Blondel2008,newmangirvan,Porter2009},
hierarchy \cite{Ravasz2003}, synchronizability \cite{Barahona2002}, degree
assortativity \cite{Newman2002}, robustness to targeted and random
attacks \cite{Albert2000}, the Rent exponent \cite{Landman1971}, and mean
connection distance \cite{Bassett2008}.

In our descriptions below, we give for each diagnostic (i) a mathematical
definition, (ii) an intuitive description of the term, and (iii) a comment on
its possible physical significance for the granular system that we study. We
also computed node-specific values for the following diagnostics: geodesic
betweenness, global efficiency, and clustering coefficient.  (See the
discussions below.)

\begin{enumerate}

\item Number of nodes $N$: (i) The diagnostic $N$ is defined as the
    number of nodes in a network. (ii) It is used as a measure of the
    size of a system. (iii) In this study, $N$ is the number of
    particles in the system.  It provides a consistent but dynamically
    uninteresting characterization of the network because it is identical
    at all points in time.

\item Number of edges $D$: (i) The diagnostic $D$ is defined as
    $D=\sum_{ij} A_{ij}$, where ${\bf A}$ is an unweighted (binary)
    network with components $A_{ij}$. Nodes are particles, and an
    edge exists between particles $i$ and $j$ (i.e, $A_{ij} = 1$) if and
    only if particles $i$ and $j$ are in contact with each other
    (otherwise, $A_{ij} = 0$). (ii) The quantity $D$ is simply the total
    number of edges in the system. (iii) The number of edges $D$ is
    related to the mean contact number, which is denoted by $z$ in the
    granular-materials community. The mean contact number of the system
    is equal to $z=\frac{D}{2N}$. The diagnostic $D$ provides a consistent
    but uninteresting characterization of the network because the number
    of contacts scales with pressure  \cite{VanHecke2010} (which is the
    same for all experimental runs).

\item Global efficiency $E$  \cite{Latora2001}: (i) Let $d_{ij}$ be the
    shortest (geodesic) number of steps necessary to get from node $i$ to
    node $j$.  The global efficiency is then defined as
    \begin{equation}
        E = \frac{1}{N(N-1)}\sum_{i \neq j}\frac{1}{d_{ij}}\,.
        \label{eq:Ewc}
    \end{equation}
    (ii) Global efficiency can be interpreted as a measure of how well a
    signal is transmitted through a network. (iii) One can expect the
    global efficiency to be small in 2D granular packings because
    particles that are not geographically close to one another are
    separated by multiple contacts (edges) and therefore by a long path
    length (low efficiency).  As one can see in Table \ref{Tab0}, this is
    indeed the case.

\item Geodesic node betweenness $B$ \cite{Freeman1977}: (i) Geodesic node
    betweenness is defined for the $i^{th}$ node in a network
    $\mathcal{G}$ as
        \begin{equation}
            B_{i} = \sum_{j,m,i \in \mathcal{G}} \frac{\psi_{j,m}(i)}{\psi_{j,m}}\,,
            \label{eq:Bwc}
        \end{equation}
where all three nodes ($j$, $m$, and $i$) must be different from each
other, $\psi_{j,m}$ is the number of geodesic paths between nodes $j$ and
$m$, and $\psi_{j,m}(i)$ is the number of geodesic paths between $j$ and
$m$ that pass through node $i$. The geodesic betweenness of an entire
network $B$ is defined as the mean of $B_{i}$ over all nodes $i$ in the
network. (ii) Geodesic betweenness can be interpreted as a measure of
traffic flow on a network. (iii) One might expect the majority of geodesic
paths that link any node of the packing to any other node to pass through
the middle of the system. Indeed, we find that the largest values of betweenness occur in the center of the system and the smallest values along the edges of the packing.

\item Random-walk node betweenness $B_{rw}$  \cite{Newman2005}: (i) For an
    adjacency matrix ${\bf A}$ and diagonal matrix ${\bf D}$, let ${\bf M}_{t}
    = {\bf A}_{t}\cdot {\bf D}_{t}^{-1}$ be the matrix ${\bf M}$ with the row and
    column $t$ removed (and ${\bf A}_{t}$ and ${\bf D}_{t}$ are defined analogously).
    The probability that a walk starts at $s$, takes $n$ steps, and ends
    up at some node $i$ (which cannot be $t$ because $t$ has been removed) is
    given by element $is$ of ${\bf M}_{t}^{n}$; denote this element by
    $[{\bf M}_{t}^{n}]_{is}$. Walks end up at $v$ and $w$ with probabilities
    $[{\bf M}_{t}^{n}]_{vs}$ and $[{\bf M}_{t}^{n}]_{ws}$.  Fractions
    $1/k_{v}$ and $1/k_{w}$ of these walks subsequently pass along the edge
    $(v,w)$ in one direction or the other, assuming that such an edge
    exists. (Note that $k_{v}$ is the degree of $v$ and $k_{w}$ is the
    degree of $w$.) Summing over all $n$ shows that the mean number
    of times that a walk of any length traverses the edge from $v$ to $w$
    is $k_{v}^{-1}[({\bf I} - {\bf M}_{t})^{-1}]_{vs}$. The random-walk betweenness
    of a node is the mean of this quantity over all edges emanating from
    that node, and the random-walk betweenness of the entire network is
    the mean of the random-walk betweenness of all nodes in the network.
    (ii) Random-walk betweenness can be interpreted as a measure of
    information flow or signal flow in a network. (iii) Similar to geodesic node betweenness, one might expect the random-walk node
    betweenness to be highest in the center of the system and lowest on
    the edges of the system.  This is indeed the case.

\item Geodesic edge betweenness $B_{e}$  \cite{Girvan2002}: (i) Inspired
    by Freeman's geodesic node betweenness, the geodesic edge betweenness of an
    edge is defined as the number of shortest paths between pairs of
    nodes that run along it. For the edge connecting nodes $j$ and $m$,
    the geodesic edge betweenness is given by
    \begin{equation}
    	B_{e}(j,m) = \sum_{i,k}\psi_{i,k}(j,m)\,,
    \end{equation}
    where $\psi_{i,k}(j,m)$ is the number of shortest paths between $i$ and
    $k$ that pass through the edge connecting nodes $j$ and $m$. (ii) One can interpret edge betweenness as a measure of the influence of
    an edge on traffic flow through a network. (iii) In a 2D granular
    packing, edge betweenness might indicate the influence of a
    contact on a hypothetical flow through the network. In our system, we find that edge betweenness is largest in the center of the system and smallest on the edges of the
    system.  This is consistent with the results for the geodesic node
    betweenness.

\item Eigenvector centrality $C_{e}$  \cite{Bonacich1972b}: (i) The
    eigenvector centrality $C_{e}(i)$ of node $i$ is proportional to the sum of the
    centralities of the nodes connected to it:
     \begin{equation}\label{evec}
     	C_{e}(i) = \frac{1}{\lambda} \sum_{j \in M(i)} C_{e}(j) = \frac{1}{\lambda} \sum_{j} A_{ij} C_{e}(j)\,,
     \end{equation}
    where $M(i)$ is the set of nodes that are neighbors of $i$ (i.e., which are connected to $i$ directly via an edge) and $\lambda$ is the largest eigenvalue of ${\bf A}$.  From equation
    (\ref{evec}), one can deduce that $C_{e}(i)$ is the $i^{th}$
    component of the leading eigenvector (each entry of which is positive
    by the Perron-Frobenius theorem \cite{newman2010}) of the adjacency
    matrix.  (ii) Eigenvector centrality can be used to measure the
    importance of a node in a network based on its direct connection to
    important nodes. (iii) In a 2D granular packing, one would expect eigenvector centrality to be large for a particle that has many
    contacts or for a particle whose immediate neighbors have many
    contacts. Indeed, we find that eigenvector
    centrality is highest in a local region of the system in which
    high-degree nodes are most concentrated.

\item Closeness centrality $C_{c}$ \cite{Sabidussi1966}: (i) We use a
    version of closeness centrality that is appropriate for both
    connected and disconnected graphs \cite{Dangalchev2006}.  It is
    defined as
    \begin{equation}\label{close}
    	C_{c}(i) = \sum_{j \in V / i} 2^{- \psi_{\mathcal{G}}(i,j)}\,,
    \end{equation}
    where $\psi_{\mathcal{G}}(i,j)$ is the geodesic distance between nodes $i$ and $j$ (i.e., the length of the shortest path connecting $i$ and $j$) and the
    notation $V/i$ indicates that $V$ is the connected network component
    reachable from $i$  and does not include $i$. (ii) Closeness centrality can be used as a
    measure of the importance of a node in a network. (iii) For 2D
    granular systems, one might expect closeness centrality to be
    small given the lattice-like topology of
    a contact network. However, as shown in Table~\ref{Tab0}, closeness
    values are somewhat larger than those for random geometric graphs
    (RGGs; see the discussion in Appendix \ref{appc}

\item Subgraph centrality $C_{s}$  \cite{Estrada2005}: (i) We first note
    that the number of closed walks of length $k$ starting and ending at
    node $i$ is given by the $k^{th}$ local spectral moment $\mu_{k}(i)$,
    which is defined as the $i^{th}$ diagonal entry of the
    $k^{th}$ power of the adjacency matrix ${\bf A}$:
    \begin{equation}
    	\mu_{k}(i) = [{\bf A}^{k}]_{ii}\,.
    \end{equation}
The subgraph centrality of node $i$ is then defined as
    \begin{equation}
 	   C_{s}(i) = \sum_{k=0}^{\infty} \frac{\mu_{k}(i)}{k!}\,.
    \end{equation}
    (ii) Subgraph centrality characterizes the participation of each node
    in all subgraphs in a network. (iii) For 2D granular systems, one
    might expect subgraph centrality to be small because nodes
    participate in few subgraphs other than their own, as their
    connectivity is strongly constrained to their local geographic
    neighborhood.  Indeed, as indicated in Table \ref{Tab0}, the subgraph
    centrality has a value that is less than 1/3 that of the value that we computed for a
    corresponding ensemble of RGGs (see our later discussion).

\item Communicability $Co$ \cite{Estrada2008}: Because of the direct
    relationship between the powers of the adjacency matrix ${\bf A}$ and
    the number of walks in a network, one can define the communicability
    between nodes $i$ and $j$ as
    \begin{equation}
        Co_{ij} = \sum_{k=0}^{\infty} \frac{[{\bf A}^{k}]_{ij}}{k!}\,.
    \end{equation}
The communicability $Co$ of a network is then the mean of the
communicabilities of each pair of (nonidentical) nodes.  (ii)
Communicability was developed to measure the ease of communication or
transmission in terms of passage between different nodes in a network,
and it is specifically based on walks rather than paths
 \cite{Estrada2008}. (iii) For 2D granular systems, one might expect the
mean communicability to be small because the geographic nature of the
contacts creates a lattice-like topology.  Indeed, as indicated in Table
\ref{Tab0}, its value is less than 1/4 than the value that we computed for a
corresponding ensemble of RGGs.

\item Clustering coefficient $C$ \cite{Watts1998}: (i) The diagnostic $C$
    is defined by supposing that a node $i$ has $k_{i}$ neighbors, so a
    maximum of $k_{i} (k_{i} - 1)/2$ edges can exist between these
    neighbors. The local clustering coefficient $C_{i}$ is the fraction
    of these possible edges that actually exist:
\begin{equation}
	C_i=\frac{\sum_{mj} A_{mj} A_{im} A_{ij}} {k_i(k_i-1)}\,. \label{eq:Cc}
\end{equation}
    The clustering coefficient $C$ of an entire network is then defined
    as the mean of $C_{i}$ over all nodes $i$. (ii) The clustering
    coefficient $C$ can be interpreted as a measure of local clustering
    properties in a network. (iii) One can expect $C$ to be large in 2D
    granular packings because particles that are geographically close to
    one another are also near each other in a network.  This
    ought to yield a large number of connected triples and hence a high
    value of $C$.  As shown in Table \ref{Tab0}, we do indeed observe
    reasonably large values \cite{newman2003} for clustering
    coefficients in the 17 experimental runs (the mean value over all
    runs is $C \approx 0.26$), but interestingly the mean value of $C$ in the
    corresponding RGG ensemble is twice as high.

\item Local efficiency $E_{l}$  \cite{Latora2001}: (i) The the local efficiency of node $i$ is defined as
\begin{equation}
        	E_{l}(i) = \frac{1}{N_{G_{i}}(N_{G_{i}}-1)} \sum_{j,k \in G_{i}} \frac{1}{d_{j,k}}\,,
\end{equation}
where $G_{i}$ is the subgraph consisting of all nodes connected to node $i$ along with
all of their edges between each other, and $d_{j,k}$ is the minimum path
length between nodes $j$ and $k$ in this subgraph.
The local efficiency $E_l$ is the mean value of $E_l(i)$ over all nodes $i$.
(ii) Local efficiency $E_{l}$ can be interpreted as a measure how well a signal is transmitted through a subgraph. (iii) One might expect local efficiency to be
large in 2D granular packings because particles that are very
close to each other geographically lie in one another's subgraphs.
However, as we show in Table \ref{Tab0}, we obtain values that are
only about half of those for corresponding RGGs.  The mean
granular-network value of $0.33$ is comparable in value to some
communication networks \cite{Latora2001}.

\item Modularity index $Q$  \cite{newmangirvan,Porter2009,Fortunato2010}:
    (i) Networks can be partitioned into communities (or modules) in
    which nodes inside the same community are more densely connected to
    each other than they are to nodes in other communities. The
    modularity of a network partition is defined as
   \begin{equation}\label{modmod}
        Q = \frac{1}{2D} \sum_{ij} \left[A_{ij} - \frac{k_{i} k_{j} }{2D}\right] \delta_{g_{i},g_{j}}\,,
    \end{equation}
    where $k_{i}$ is the degree of node $i$, $D$ is the total number of
    edges in the network, $\delta_{ij}$ is the Kronecker delta,
    and $g_{i}$ is the community to which node $i$ has been assigned.
    With the standard null model $P_{ij} = {k_{i} k_{j} }/{(2D)}$,
    equation (\ref{modmod}) is sometimes called `Newman-Girvan modularity.'  One
    uses one of numerous possible computational heuristics to maximize
    $Q$ in the space of all network partitions, and one can then report
    the maximum value obtained for $Q$. However, it is important to note
    that the optimization of $Q$ is NP-hard  \cite{brandes08}, so one
    cannot expect the output of an optimization algorithm to be a
    globally optimal partition of a network. In this light,
    we use two different computational heuristics to optimize
    $Q$: Newman's spectral algorithm  \cite{Newman2006} (which yields a
    modularity value that we denote $Q_{s}$) and the Louvain locally
    greedy method  \cite{Blondel2008} (yielding a modularity value that we denote
    $Q_{L}$). (ii) The optimal value of $Q$ is a measure of how well a
    network can be partitioned into cohesive communities. (iii) In a 2D
    granular system, one might expect communities to be localized
    geographically because connectivity between nodes in potential
    communities is constrained geographically.  Indeed, as shown in Table
    \ref{Tab0}, the values of $Q_s$ and $Q_L$ are both extremely high
     \cite{Fortunato2010,Porter2009}.

\item Hierarchy $h$  \cite{Ravasz2003}: (i) A sense of hierarchical
    structure in a network can be characterized by the coefficient $h$,
    which is used to quantify a putative power-law relationship between clustering coefficient $C_i$ and the degree $k_i$ of all nodes in
    the network  \cite{Ravasz2003}:
    \begin{equation}\label{hierarchy}
        C_i \sim k_i^{- h}\,.
    \end{equation}
    (ii) Networks in which clustering coefficient has a power-law scaling
    with degree possess a hierarchy in which hubs (i.e., high-degree
    nodes) tend to have low local clustering and low-degree nodes tend to
    have high local clustering.  The parameter $h$ gives a precise
    scaling of such effects when the relation (\ref{hierarchy}) holds,
    and it can perhaps indicate a looser sense of hierarchy in more
    general situations.  (iii) It is not clear a priori whether 2D
    granular packings have some hierarchical characteristics,
    though the authors of Ref.~ \cite{Ravasz2003} have suggested that
    geographic networks are not very hierarchical.  Our calculations (see
    Table \ref{Tab0}) provide some support for this claim, as we observe
    significant scaling between $C_i$ and $k_i$ and obtain $h \approx
    0.76$.  It is noteworthy, however, that this value is roughly three
    times what one obtains in a corresponding RGG ensemble (see Table
    \ref{Tab0}).

\item Synchronizability $s$  \cite{Barahona2002}: (i) The
    synchronizability is defined as
    \begin{equation}
        	s = \frac{\lambda_{2}}{\lambda_{N}}\,,
    \end{equation}
    where $\lambda_{2}$ is the second smallest eigenvalue of the
    Laplacian $\mathcal{L}$ of the adjacency matrix and $\lambda_{N}$ is
    the largest eigenvalue of $\mathcal{L}$  \cite{Barahona2002}. (ii) The
    synchronizability of a network characterizes structural properties of
    a network that hypothetically enable it to synchronize rapidly. (iii)
    One might expect that the synchronizability of the contact network in
    a 2D granular packing is small due to the lattice-like nature
    of the network topology.  Indeed, as shown in Table \ref{Tab0}, the
    value for $s$ for our system is tiny.

\item Degree assortativity $a$ \cite{Newman2002}: (i) The degree
    assortativity of a network (which is often called simply
    `assortativity') is defined as
    \begin{equation}
      a= \frac{E^{-1} \sum_{i} j_{i} k_{i} - \left[E^{-1} \sum_{i} \frac{1}{2} (j_{i} + k_{i})\right]^{2} } {E^{-1} \sum_{i} \frac{1}{2} (j_{i}^{2} + k_{i}^{2}) - \left[E^{-1} \sum_{i} \frac{1}{2} (j_{i}+k_{i})\right]^{2}}\,,
    \end{equation}
    where $j_{i}$ and $k_{i}$ are the degrees of the nodes at the two ends of the
    $i^{th}$ edge ($i \in \{1\,,\ldots\,, E\}$  \cite{Newman2006}. (ii)
    Degree assortativity measures the preference of a node to connect to other
    nodes of similar degree (leading to an assortative network, for which
    $a>0$) or to nodes of very different degree (leading to a
    disassortative network, for which $a<0$). (iii) It is not clear a
    priori whether 2D granular packings should display degree
    assortativity.  Our calculations indicate that the degrees exhibit
    some mild positive assortativity ($a \approx 0.14$), but the
    corresponding RGG ensembles have a significantly higher positive
    assortativity of $a \approx 0.56$.  (See Table \ref{Tab0}.)

\item Robustness $R$ \cite{Albert2000}: (i) One can define robustness in
    terms of different types of attacks on a network. In the most commonly
    studied type of targeted attack, nodes are removed (one by one) in
    descending order of their degree; in a random attack, nodes are
    removed in random order. Each time a node is removed from a
    network, we recalculate the size $S$ (i.e., number of nodes) of the largest connected component.  One can examine robustness by plotting $S$ versus the number of nodes
    removed $n$ \cite{Achard2006,Lynall2010,Bassett2010c}. One can then define a robustness
    parameter $R$ as the area under the curve in the plot of $S = S(n)$. More robust networks retain a larger connected component
    even when several nodes have been removed; this is represented by
    a larger area under the curve and hence by larger values of $R$. (ii)
    Robustness indicates network resilience to either targeted
    ($R_{t}$) or random ($R_{r}$) attacks. (iii) Robustness tends to be
    most interesting for networks with highly heterogeneous degree
    distributions, so it might not be very insightful for 2D granular
    packings.  We note, however, that we find values of $R_t$ and $R_r$ for our
    granular networks that are more than 3 times as
    large as those for the corresponding RGG ensemble.  (See Table \ref{Tab0}.)

\item Rent exponent $p$  \cite{Christie2000}: (i) Rent's rule, which was
    first discovered in relation to computer chip design, defines a
    scaling relationship between the number of external signal
    connections (edges) $e$ to a block of logic and the number of
    connected nodes $n$ in the block \cite{Christie2000}:
    \begin{equation}
         e\sim n^p\,,
    \end{equation}
    where $p \in [0,1]$ is the Rent exponent. (ii) The Rent exponent
    measures the efficiency of the physical embedding of a topological
    structure. (iii) Due to the physical constraints of a 2D granular packing, we expect to observe Rentian scaling with a relatively low
    Rent exponent (similar to that of a lattice). The theoretically expected minimum of the Rent's exponent for a 2-dimensional physical lattice is $p_{t}=1-1/D_{E}$ \cite{Ozaktas1992,Bassett2010,Verplaetse2001} where $D_{E}$ is the Euclidean dimension of the space (e.g., 2), and therefore $p_{t}=0.5$, which is consistent with empirical results on memory circuits \cite{Russo1972}. However, the expected value of the Rent exponent might further depend on the type of physical lattice under study (e.g., a rectangular or hexagonal lattice).

\item Mean connection distance $mcd$: (i) An edge's estimated connection
    distance $L_{ij}$ is defined as the Euclidean distance between the
    centroids of particles $i$ and $j$. (ii) The mean connection distance
    $mcd$ is defined as the mean of all $L_{ij}$ values in the network.
    (iii) The mean connection distance in a 2D granular packing is
    related to the number of particles, the area of the system, and
    particle size.

\end{enumerate}


\subsection*{Diagnostics Applied to Force-Weighted Contact Networks}

To characterize the structure of the force-weighted networks, we used 8
diagnostics: normalized strength \cite{horvath2011,Barrat2004},
diversity \cite{Campbell1986}, path length \cite{Dijkstra1959}, geodesic node
betweenness \cite{Freeman1977,Barrat2004}, geodesic edge
betweenness \cite{Girvan2002}, clustering coefficient \cite{Barrat2004},
transitivity \cite{Onnela2005}, and optimized modularity \cite{newmangirvan}.

\begin{enumerate}

\item Normalized Strength $S'$  \cite{horvath2011,Barrat2004}: (i) The
    strength of node $i$ is given by the column sum of the weighted
    adjacency matrix:
    \begin{equation}
    	S_i = \sum_{j} W_{ij}\,,
    \end{equation}
    and the strength of an entire weighted network ${\bf W}$ is the mean
    of $S_{i}$ over all $i$. The normalized strength $S'$ is
    \begin{equation}
    	S'_i = \frac{S_{i}}{N}\,,
    \end{equation}		
    where $N$ is the total number of nodes. (ii) Strength is a measure of how strong the
    connections are in a network. (iii) In the present context,
    normalized strength provides a measure of the mean contact forces between
    particles, and we therefore expect this diagnostic to be correlated
    with sound propagation.  Indeed, we observe this in our calculations.

\item Diversity $V$ \cite{Campbell1986}: (i) The diversity of node $i$ is
    defined as the variance of the edge weights for the set of all edges
    connected that are connected to it.  It is given by
    \begin{equation}
        V_{i} = \left[\frac{1}{N} \sum_{j} \left(W_{ij}-\langle W_i \rangle\right)^{2}\right]^{1/2}\,.
    \end{equation}
    The diversity of an entire weighted network is the mean of
    $V_{i}$ over all $i$. (ii) Diversity is a measure of the variance of
    connectivity strengths in a network. (iii) In the present context,
    diversity is a measure of the variance of contact forces between
    particles, and it has a high positive correlation with
    normalized strength.

\item Global efficiency $E_{w}$  \cite{Latora2001}: (i) Let $d^w_{ij}=\mathrm{max}(W_{ij})-W_{ij}$ be the weighted shortest path between nodes $i$ and $j$.
    The global efficiency of node $i$ is then defined as
    \begin{equation}
        	E_{w}(i)=\frac{1}{N-1}\sum_{j \neq i}\frac{1}{d^w_{ij}}\,.
	   \label{eq:Eww}
    \end{equation}
    The global efficiency $E_{w}$ is the mean value of $E_w(i)$ over all nodes $i$. (ii) One can interpret global efficiency as a measure of how
    efficiently a signal is transmitted through a network. (iii) We
    expect the global efficiency of the force-weighted contact network to
    be large in the center of the packing and small on the edges of the
    packing because particles that are not geographically close to each
    other do not exert forces on one another.  Indeed, this is what we
    observe.

\item Clustering coefficient $C_{w}$  \cite{Barrat2004}: (i) One can
    define a weighted clustering coefficient $C_{w}(i)$ of node $i$ using
    the formula
     \begin{equation}
     		C_{w}(i) = \frac{1}{S_{i}(k_{i}-1)} \sum_{j,k} \frac{(W_{ij} + W_{ik})}{2} A_{ij} A_{ik} A_{jk}\,,
		\label{eq:Cw}
     \end{equation}
    where $S_{i}$ is node $i$'s strength, $k_{i}$ is its degree, ${\bf W}$ is the weighted adjacency
    matrix, and ${\bf A}$ is the underlying binary adjacency matrix. (ii)
    The weighted clustering coefficient $C_w(i)$ measures the strength of local
    connectivity. (iii) It is constrained by the underlying contact network structure, so we
    expect it to have a high positive correlation with the binary clustering
    coefficient $C(i)$. Indeed, the Pearson correlation coefficient between the binary and
    weighted clustering coefficients over the experimental runs is $r \approx 0.94$
    (with a p-value of $p \approx 2\times 10^{-9}$).  Both diagnostics tend
    to attain their highest values on the edges of the packing, where
    nodes' immediate neighbors are most likely to also be connected to
    one another.

\item Geodesic node betweenness $B_{w}$  \cite{Park2004}: (i) Geodesic
    betweenness is defined for the $i^{th}$ node in a network
    $\mathcal{G}$ as
        \begin{equation}
            B_{w}(i) = \sum_{j, m, i \in \mathcal{G}} \frac{\tilde{\psi}_{j,m}(i)}{\tilde{\psi}_{j,m}}\,,
            \label{eq:Bww}
        \end{equation}
    where all three nodes ($j$, $m$, and $i$) must be different from each
    other, $\tilde{\psi}_{j,m}$ denotes the number of geodesic weighted
    paths between nodes $j$ and $m$, and $\tilde{\psi}_{j,m}(i)$ denotes
    the number of geodesic weighted paths between $j$ and $m$ that pass
    through node $i$.
(As with weighted global efficiency, the weighted shortest path between nodes $i$ and $j$ is $d^w_{ij}$.)
 The weighted geodesic betweenness of an entire
    network $B_{w}$ is defined as the mean of $B_{w}(i)$ over all nodes
    $i$. (ii) One can interpret weighted geodesic betweenness as a
    measure of traffic flow on a network. (iii)  We expect
    betweenness in weighted networks to correlate positively with
    strength, just as betweenness in binary networks correlates
    positively with degree \cite{Newman2005}. In a 2D granular packing,
    we expect particles in the center of the system to
    have high values of weighted betweenness because more paths must pass
    through them to connect the edges of the system.  We indeed find this to
    be the case.

\item Geodesic edge betweenness $B_{ew}$ \cite{Newman2002,Park2004}: (i)
    We define geodesic edge betweenness on weighted networks using the
    number of shortest weighted paths between pairs of nodes that run
    along it.  (We again determine the path distance between nodes $i$ and $j$ using $d^w_{ij}$.) For the edge connecting nodes $j$ and $m$, the weighted
    geodesic edge betweenness is therefore
   \begin{equation}
 	   B_{ew}(j,m) = \sum_{i,k}\tilde{\psi}_{i,k}(j,m)\,,
    \end{equation}
    where $\tilde{\psi}_{i,k}(j,m)$ is the number of shortest paths
    between nodes $i$ and $k$ that pass through the edge connecting nodes
    $j$ and $m$. (ii) Weighted edge betweenness indicates the influence
    of an edge on traffic flow through a network. (iii) In a 2D granular
    packing, the edge betweenness should give an indication of the influence
    of a contact on a hypothetical flow through the network. We find that edge betweenness is largest in the center
    of the system because more paths must pass through these edges to
    connect all pairs of particles.

\item Modularity index $Q_{w}$
     \cite{newmangirvan,Porter2009,Fortunato2010}: The weighted modularity
    of a network partition is
    \begin{equation} \label{eq:Xw}
        Q_{w} = \frac{1}{2\bar{W}} \sum_{ij} \left[W_{ij} - \frac{S_{i} S_{j} }{2\bar{W}}\right] \delta_{g_{i},g_{j}}\,,
    \end{equation}
    where $S_{i}$ is node $i$'s strength, $\bar{W}$ is the total
    strength of the edges in a network, $W_{ij}$ is an element of the
    weighted adjacency matrix, $\delta_{ij}$ is the Kronecker delta, and $g_{i}$ is the label of the community to which node $i$
    has been assigned.  As with unweighted networks, one uses some
    computational heuristic to find a partition that maximizes $Q$. As with the binary networks, we have used the Louvain locally greedy optimization
    method \cite{Blondel2008}.
    (ii) The maximum value of $Q$ is a measure
    of how well a network can be partitioned into cohesive communities.
    (iii) In a 2D granular packing, in which forces between particles are
    represented as edge weights, we expect communities to be
    localized in space because the forces between nodes in
    potential communities are constrained geographically.  Indeed, as
    shown in Fig.~\ref{fig:scale}, this is indeed the case.

\item Transitivity $T$  \cite{Onnela2005}: (i) The weighted transitivity
    $T(i)$ of node $i$ is
    \begin{equation}
        	T(i) = \frac{2}{k_{i}(k_{i}+1)} \sum_{j,k} \left(\tilde{W}_{ij} \tilde{W}_{jk} \tilde{W}_{ik}\right)^{1/3}\,,
    \end{equation}
    where we have normalized weights by the maximum edge weight in the matrix:
    \begin{equation}
    	\tilde{W}_{ij} = \frac{W_{ij}}{\mathrm{max} \left(W_{ij}\right)}\,.
    \end{equation}
    The transitivity $T$ of the entire network is the mean of
    $T(i)$ over all nodes $i$. (ii) Weighted transitivity is a
    generalization of the local clustering coefficient in unweighted
    networks in which one computes the sum of the weights of edges in a
    network's triangles instead of computing simply the number of
    triangles. (iii) We expect weighted transitivity to be similar
    to the weighted clustering coefficient $C_w$. Indeed, we find that the two variables are
    highly correlated over experimental runs (with a Pearson correlation coefficient of $r \approx 0.99$ and a p-value of $p \approx 1 \times 10^{-16}$).

\end{enumerate}

We implemented all computational and simple statistical operations (such as t-tests and
correlations) using MATLAB$^{\textregistered}$ (2009a, The
MathWorks Inc., Natick, MA). We estimated network diagnostics using
a combination of in-house software, the Brain Connectivity Toolbox
 \cite{Rubinov2009}, the MATLAB Boost Graph Library, and the fast unfolding
community detection code for the Louvain optimization of modularity
 \cite{Blondel2008} from Peter Mucha  \cite{Jutla2011}.


\section{Appendix B: Reliability of Network Structure}

\begin{figure}
\begin{center}
\includegraphics[width=.35\textwidth]{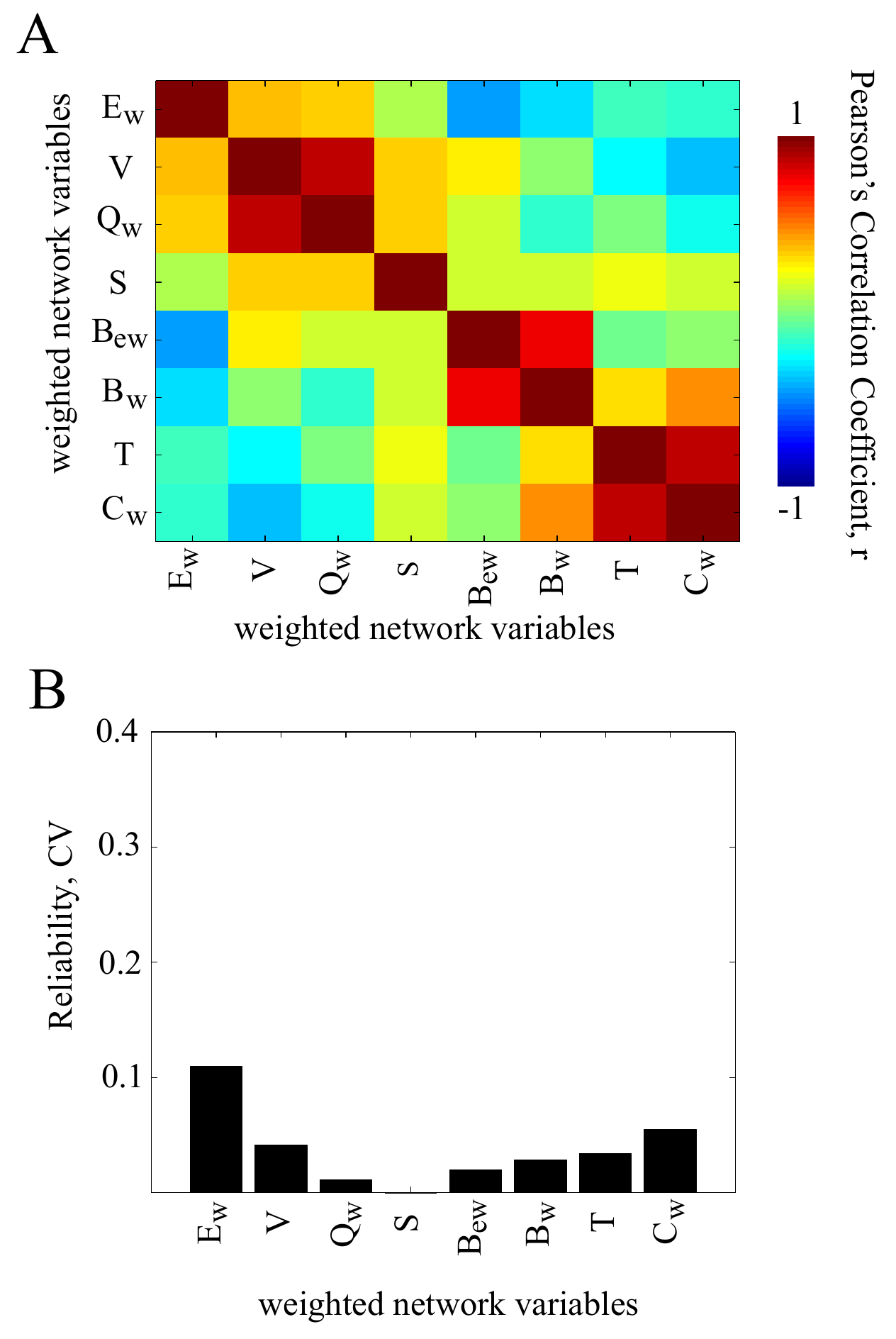}
\caption{ [Color online]
(A) Relationships between 8 diagnostics applied to the weighted
networks: geodesic node betweenness ($B_{w}$), clustering coefficient
($C_{w}$), diversity ($V$), geodesic edge betweenness ($B_{ew}$), global
efficiency ($E_{w}$), modularity ($Q_{w}$) optimized using the Louvain
method, and normalized strength ($S'$). We ordered the diagnostics to maximize the correlation along the diagonal for better visualization of
highly correlated groups. Color indicates the correlation between global
network diagnostics over the 17 experimental runs. (B) Reliability, as
measured by the coefficient of variation (CV), over the 17 runs for the 8
weighted graph diagnostics reported in (A).
\label{fig:reliability2}}
\end{center}
\end{figure}

We quantify the reliability of each diagnostic by
calculating the coefficient of variation (a normalized measure of dispersion)
over the 17 experimental runs: $\mathrm{CV} = {\sigma}/{|\mu|}$, where
$\sigma$ is the standard deviation and $\mu$ is the mean. Values of
$\mathrm{CV} \lesssim 0.2$ are commonly considered to be acceptable, as they
indicate that a variable is reliable
 \cite{Bassett2010c,Deuker2009,Mancl2004}. See Table \ref{Tab1} for
$\mathrm{CV}$ values for all binary network diagnostics and
Fig.~\ref{fig:reliability2}B for a corresponding bar graph.
Interestingly, reliable diagnostics are dispersed among the quantities we considered rather than focused on sets of related diagnostics.

\begin{figure}
\begin{center}
\includegraphics[width=.35\textwidth]{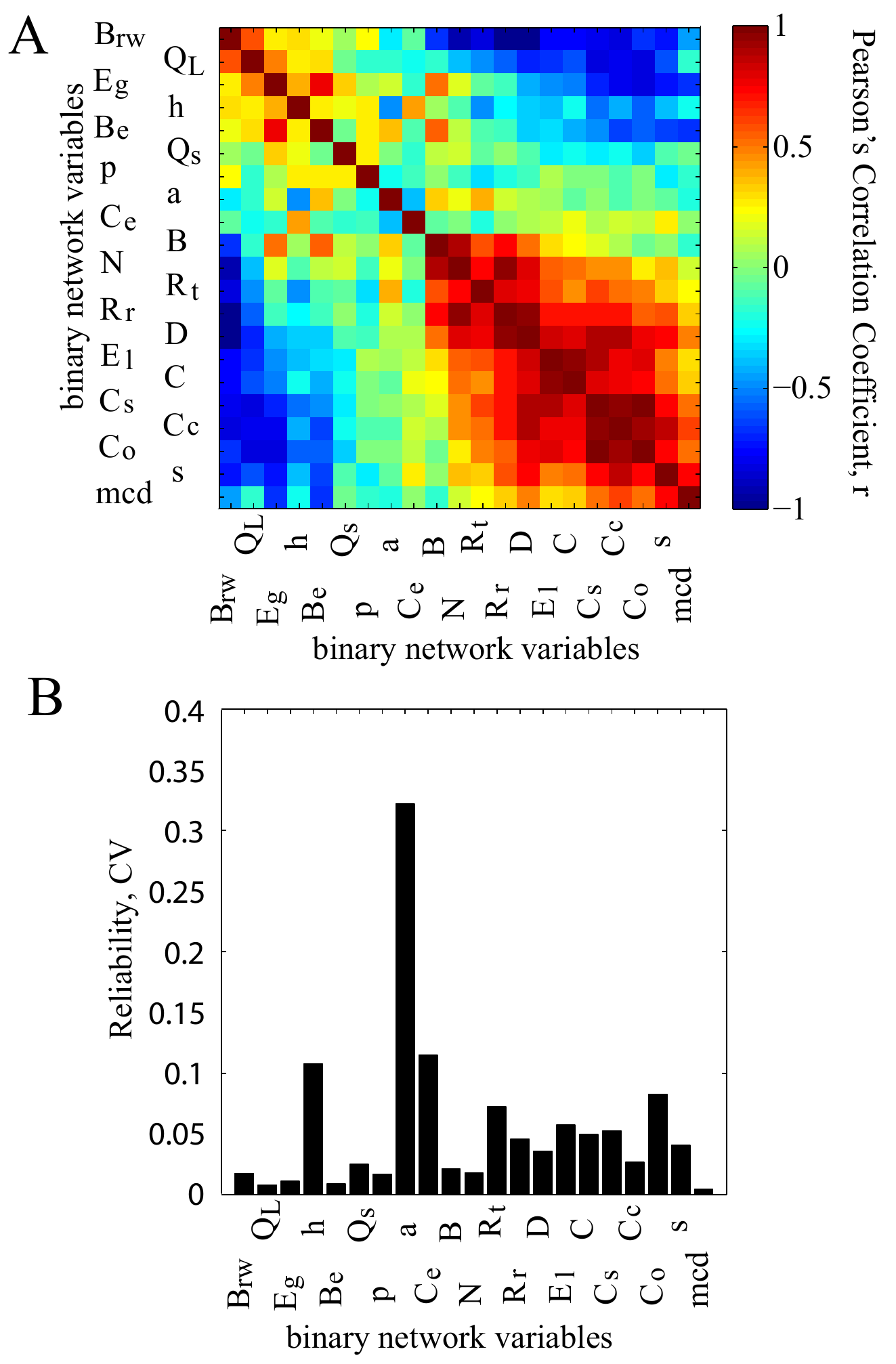}
\caption{ [Color online]
(A) Relationships between 21 binary network diagnostics: degree assortativity ($a$), geodesic node betweenness ($B$), closeness centrality ($C_{c}$), clustering coefficient
($C$), communicability ($Co$), geodesic edge betweenness ($B_{e}$),
eigenvector centrality ($C_{e}$), global efficiency ($E_{g}$), hierarchy
($h$), local efficiency ($E_{l}$), modularity optimized using the Louvain
($Q_{L}$) and spectral ($Q_{s}$) heuristics, number of nodes ($N$), number of
edges ($D$), mean connection distance ($mcd$), random-walk node betweenness
($B_{rw}$), Rent exponent ($p$), robustness to random attack ($R_{r}$),
robustness to targeted attack ($R_{t}$), subgraph centrality ($C_{s}$), and
synchronizability ($s$). We order the diagnostics to maximize the correlation
along the diagonal for better visualization of highly correlated groups of diagnostics.
Color indicates the correlation between global network diagnostics over the
17 experimental runs.
(B) Reliability, as measured by the coefficient
of variation (CV), over the 17 runs for the 21 binary network diagnostics
reported in (A).
 \label{fig:reliability1}
}
\end{center}
\end{figure}

\begin{table}
\begin{center}
\begin{tabular}{| l | l | l |}
\hline
Binary Contact Network Diagnostic & Variable & CV \\
\hline \hline
Mean Connection Distance & $mcd$ 	& 0.0046 \\
Geodesic Edge Betweenness & $B_{e}$ 	& 0.0090 \\
{\bf Global Efficiency}        & $E_{g}$ 	& 0.0114 \\
Rent Exponent          & $p$		& 0.0171 \\
Random-Walk Node Betweenness  & $B_{rw}$ & 0.0176 \\
Number of Nodes          & $N$		& 0.0179 \\
{\bf Geodesic Node Betweenness} & $B$		& 0.0213 \\
{\bf Modularity: Louvain Optimization}       & $Q_{L}$	& 0.0071 \\
Modularity: Spectral Optimization        & $Q_{s}$	& 0.0252 \\
Closeness Centrality     & $C_{c}$ 	& 0.0267 \\
Number of Edges          & $D$		& 0.0360 \\
Synchronizability        & $s$		& 0.0408 \\
Robustness, Random       & $R_{r}$	& 0.0456 \\
{\bf Clustering Coefficient}   & $C$		& 0.0496 \\
Subgraph Centrality      & $C_{s}$	& 0.0528 \\
Local Efficiency         & $E_{l}$	& 0.0573 \\
Robustness, Targeted     & $R_{t}$	& 0.0727 \\
Communicability          & $Co$		& 0.0829 \\
Hierarchy                & $h$		& 0.1080 \\
Eigenvector Centrality   & $C_{e}$	& 0.1150 \\
Degree Assortativity            & $a$		& 0.3219 \\
\hline
\end{tabular}
\end{center}
\caption{Reliability of network diagnostics for the binary contact networks. We measure reliability using coefficient of variance (CV).
\label{Tab1}}
\end{table}

For the force-weighted granular networks, we find lower reliability (i.e., higher values of CV) for the diagnostics than for the (binary) contact networks. Compare Fig.~\ref{fig:reliability1}B to Fig.~\ref{fig:reliability2}B and Table~\ref{Tab2} to Table~\ref{Tab1}). It is possible that the reliability is lower in the weighted networks because a large ensemble of force-chain networks are consistent with a given packing \cite{Snoeijer-2004-FNE}. Therefore, for each binary network, we are sampling one weighted network out of the many possible force-chain networks that could arise from the underlying contacts. Based on this degeneracy, we might expect that network diagnostics that depend on the force topology might be less consistent across experiments than those based on contacts alone. As with the unweighted networks, the strongly reliable weighted-network diagnostics are dispersed among the 8 diagnostics rather than focused on sets of related quantities.

\begin{table}
\begin{center}
\begin{tabular}{| l | l | l |}
\hline
Force-Weighted Contact Network Diagnostic & Variable & CV \\
\hline \hline
Transitivity             & $T$		& 0.0339 \\
{\bf Clustering Coefficient}	 & $C_{w}$	& 0.0549 \\
Geodesic Node Betweenness 	         & $B_{w}$	& 0.0282 \\
{\bf Geodesic Edge Betweenness}  	 & $B_{ew}$ 	& 0.0199 \\
Normalized Strength	         & $S'$ 		& 0.0000 \\
{\bf Modularity: Louvain Optimization} & $Q_{w}$	& 0.0135 \\
Diversity	         & $V$		& 0.0412 \\
{\bf Global Efficiency}        & $E_{w}$  & 0.1094 \\
\hline
\end{tabular}
\end{center}
\caption{Reliability of network diagnostics for the force-weighted contact networks. We measure reliability using coefficient of variance (CV).}
\label{Tab2}
\end{table}


\section{Appendix C: Comparison of Contact Networks to Random Geometric Graphs} \label{appc}

\begin{table*}
\begin{center}
\begin{tabular}{| l | l | l | l | l |}
\hline
Diagnostic Name & $EG$ & $RGG$ & $t$ & $p$ \\
\hline \hline
$EG$ Less Than $RGG$ & ~ & ~ & ~ & ~\\
\hline
Local Efficiency          & 0.33  &  0.67 & 73.33   & $3.1\times 10^{-37}$ \\
Modularity: Louvain Optimization        & 0.81  & 0.92  & 70.52   & $7.2\times 10^{-36}$ \\
Clustering Coefficient    & 0.26  &  0.54 & 85.28   & $2.5\times 10^{-30}$ \\
Random-Walk Betweenness   & 0.09  & 0.20  & 39.73   & $8.2\times 10^{-29}$ \\
Degree Assortativity             & 0.14  & 0.57  & 38.52   & $2.19\times 10^{-28}$ \\
Subgraph Centrality       & 7.19  & 28.79 &  22.41   & $3.9\times 10^{-21}$ \\
Communicability           & 0.15  & 0.73  &  17.79   & $3.6\times 10^{-18}$ \\
Rent Exponent           & 0.47  &  0.50 &  15.41   & $2.2\times 10^{-16}$ \\
\hline \hline
$EG$ Greater Than $RGG$ & ~ & ~ & ~ & ~\\
\hline
Global Efficiency         & 0.10  & 0.03 & $94.41$   & $1.0\times 10^{-40}$ \\
Mean Connection Distance  & 39.60 & 32.32 & $79.38$  & $2.5\times 10^{-38}$ \\
Robustness, Random        & $8.38\times 10^{4}$ &  $2.33\times 10^{4}$ & 54.89 & $3.0\times 10^{-33}$ \\
Closeness Centrality      & 7.84  & 4.47  & $50.29$  & $4.9\times 10^{-32}$ \\
Synchronizability         & 0.0014 & 0.0004 & $44.71$ & $2.2\times 10^{-30}$ \\
Robustness, Targeted      & $7.08\times 10^{4}$  &  $1.96\times 10^{4}$ & 36.96 & $7.9\times 10^{-28}$ \\
Edge Betweenness          & 13.33 & 4.55  & $34.71$   & $5.6\times 10^{-27}$ \\
Geodesic Node Betweenness & $6.24\times 10^{3}$  &  $2.16\times 10^{3}$ & 30.95 & $2.0\times 10^{-25}$ \\
Hierarchy                 & 0.76  &  0.27 & $24.42$  & $2.9\times 10^{-22}$ \\
Eigenvector Centrality    & 0.0172 & $0.0080$ & $19.15$ & $4.2\times 10^{-19}$ \\
Modularity: Spectral Optimization) & 0.78  &  0.75 & $5.06$   & $1.63\times 10^{-5}$ \\
\hline
\end{tabular}
\end{center}
\caption{Comparison of binary network diagnostics in the (real) experimental graphs (EGs) and the random geometric graphs (RGGs). We show
the mean values of the EGs (column 1), the mean values of the
RGGs (column 2), the t-values (column 3), and p-values
(column 4) for a two-sample t-test between the network-diagnostic values of
the two families of networks (EGs and RGGs). \label{Tab0}}
\end{table*}

Many of the diagnostics that we compute for the granular networks are highly correlated with one another (see Fig.~\ref{fig:reliability2}A).  In the (binary) contact networks, they form
roughly two families, where the correlations among the diagnostics in a given family are high. 
The diagnostics that we used for the weighted networks also exhibit some
correlations (see Fig.~\ref{fig:reliability1}A), and (unsurprisingly) this is particularly evident for diagnostics that are known to be closely related mathematically.  For example, the weighted clustering coefficient is highly correlated with transitivity, and geodesic node betweenness is highly correlated with geodesic edge betweenness.

It is important to think about the possible origins of correlations between the various network diagnostics. Some of them might be specific features of the precise granular system under consideration, but others might arise because our granular packings are confined in 2D rather than in 3D or because of our particular preparation protocol.  Still others might be general properties of spatially-embedded systems (in any dimension), of granular materials, or of networks in general.

To examine such issues, it is desirable to compare network diagnostics computed for the networks obtained from experimental data with those obtained from appropriate ensembles of null-model networks. It is common to compare the structures of networks under study to those that would be expected in Erd\"{o}s-R\'{e}nyi random graphs or from some other random graph ensemble \cite{newman2010}. The networks that we study in the present paper --- namely, contact networks in granular packings --- are spatially embedded (in the plane) because of physical constraints. The development of null-models is a wide open area of research for spatially-embedded networks \cite{Barthelemy2010}, but we can make some progress for the binary contact networks by comparing the network diagnostics in those networks to computations of the same diagnostics using an ensemble of random geometric graphs (RGGs). As we will now discuss, we find that all diagnostics (except for the ones that we fix when defining the RGG ensemble to match their counterparts in the real networks) are significantly different in the real versus random networks.

The simplest RGG  \cite{penrosergg,Dall2002,Barthelemy2010} contains $N$ nodes that are randomly and distributed according to some probability distribution throughout an ambient space, which in our case is $\mathbb{R}$ \nocite{McDiarmid2005} \footnote{Note that a `random geometric graph' in 2D is a different mathematical object from what is known as a `random planar graph' \cite{McDiarmid2005}.}.
One then places an edge between
any pair of nodes $i$ and $j$ that are separated by a distance of at most
$2r$, where one should think of the parameter $r$ as the radius of a ball
(using some metric) in the confining space.  In planar Euclidean space, one
considers a disk in $\mathbb{R}^2$ and uses ordinary (Euclidean) distance.

To compare the networks that we study to RGGs, we generated RGGs in which
we placed nodes randomly and uniformly within the 2D space of the granular
packing. For each experimental run, we created an ensemble of 100 RGGs in which the number of nodes was identical to that in the experimental
system. We likewise fix the number of edges in each RGG to be identical to
that in the real system ($D$) by choosing the threshold $2r$ so that the
number of inter-node distances less than $2r$ is equal to $D$. We calculate
the other 19 binary diagnostics (i.e., except for the number of
nodes and the number of edges, as those have been fixed to be equal in the
two sets of networks) and computed their mean over the 100 RGGs in each ensemble. By performing these computations for each experimental run, we created 1 estimate of each of the 19 diagnostic values for each of the 17 runs. We report in Table \ref{Tab0} the mean values for both
the real networks and the networks generated from the RGG ensembles.  We also
report t-values and p-values for two-sample t-tests between the values in the 17 real networks and the 17 mean values in the RGG networks. As we show in Table \ref{Tab0}, each of the 19 network
diagnostics is significantly different between the two groups. Measures of
local connectivity (e.g., clustering coefficient) are higher in the RGG, whereas
measures of global connectivity and physical distance are lower. These
results illustrate that the networks in the RGG ensemble have more locally connectivity structures than those in the 2D granular system under study.

\bibliography{bibfile3,ked5,eto3}

\end{document}